\documentclass[superscriptaddress,preprintnumbers,amssymb,nofootinbib]{revtex4}
\pdfoutput=1

\usepackage{color}
\usepackage{amsmath}
\usepackage{amsfonts}
\usepackage{amssymb}
\usepackage{array}
\usepackage{graphicx}
\usepackage{pifont}
\usepackage{phonetic}
\usepackage{bbding}
\usepackage{eufrak}
\usepackage{stmaryrd}
\usepackage{multirow}
\usepackage{booktabs}
\usepackage{ulem}
\usepackage{slashed}
\usepackage{rotating}
\usepackage{hyperref}
\usepackage{tablefootnote}
\usepackage[section]{placeins}

\renewcommand{\arraystretch}{1.5}

\usepackage{bbm}                                                  
\newcommand{\nc}{\newcommand}
\nc{\beq}{\begin{equation}}  \nc{\eeq}{\end{equation}}
\nc{\bea}{\begin{eqnarray}}  \nc{\eea}{\end{eqnarray}}
\nc{\baa}{\begin{array}}     \nc{\eaa}{\end{array}}
\nc{\bit}{\begin{itemize}}   \nc{\eit}{\end{itemize}}
\nc{\ben}{\begin{enumerate}} \nc{\een}{\end{enumerate}}
\nc{\bce}{\begin{center}}    \nc{\ece}{\end{center}}
\nc{\bpm}{\begin{pmatrix}}   \nc{\epm}{\end{pmatrix}}
\nc{\bvt}{\begin{verbatim}}  \nc{\evt}{\end{verbatim}}
\def\half{\frac12}	%\def\half{{1\over2}}

\def\to{\rightarrow}

\def\gesim{\gtrsim}

\def\boldoverdot{\,{\raise6pt\hbox{\bf.}\!\!\!\!\>}}

\def\lcal{{\cal L}}

\def\ocal{{\cal O}}

\def\pp{{\bf p}}

\def\WW{{\bf W}}

\def\mati{{\mathbbm1}}

\usepackage{bm}
\def\taubf{{\bm\tau}}		%	\def\taubf{{\pmb{$\tau$}}}
\def\ssb{spontaneous symmetry breaking}
\def\vev{vacuum expectation value}

\def\rhs{right hand side\ }

\def\diag{\hbox{\diag}}

\def\mev{\hbox{MeV}}
\def\gev{\hbox{GeV}}
\def\tev{\hbox{TeV}}

\def\vevof#1{\left\langle #1 \right\rangle}

\def\doubleundertext#1{
{\undertext{\vphantom{y}#1}}\par\nobreak\vskip-\the\baselineskip\vskip4pt%
\undertext{\hbox to 2in{}}}
\def\inbox#1{\vbox{\hrule\hbox{\vrule\kern5pt
     \vbox{\kern5pt#1\kern5pt}\kern5pt\vrule}\hrule}}
\def\sqr#1#2{{\vcenter{\hrule height.#2pt
      \hbox{\vrule width.#2pt height#1pt \kern#1pt
         \vrule width.#2pt}
      \hrule height.#2pt}}}
\def\square{\mathchoice\sqr56\sqr56\sqr{2.1}3\sqr{1.5}3}
\def\today{\ifcase\month\or
  January\or February\or March\or April\or May\or June\or
  July\or August\or September\or October\or November\or December\fi
  \space\number\day, \number\year}
\def\pmb#1{\setbox0=\hbox{#1}%
  \kern-.025em\copy0\kern-\wd0
  \kern.05em\copy0\kern-\wd0
  \kern-.025em\raise.0433em\box0 }

\def\pmbb#1{\setbox0=\hbox{#1}%
  \kern-.02em\copy0\kern-\wd0
  \kern.04em\copy0\kern-\wd0
  \kern-.02em\raise.03464em\box0 }
\def\up#1{^{\left( #1 \right) }}
\def\inv#1{\frac1{#1}}
\def\su#1{{SU(#1)}}
\def\ui{U(1)}
\def\sumprime_#1{\setbox0=\hbox{$\scriptstyle{#1}$}
  \setbox2=\hbox{$\displaystyle{\sum}$}
  \setbox4=\hbox{${}'\mathsurround=0pt$}
  \dimen0=.5\wd0 \advance\dimen0 by-.5\wd2
  \ifdim\dimen0>0pt
  \ifdim\dimen0>\wd4 \kern\wd4 \else\kern\dimen0\fi\fi
\mathop{{\sum}'}_{\kern-\wd4 #1}}
\def\nsm{$\nu$SM}
\def\el{effective Lagrangian}
\def\seff{$S_{\rm eff}$}
\def\leff{\lcal_{\rm eff}}
\def\rh{right-handed}
\def\lh{left-handed}

\def\vp{\varphi}
\def\mn{{\mu\nu}}
\def\eps{\epsilon}

\def\phit{\tilde\phi}
\def\ptd{\phit^\dagger}
\def\ncb{\overline{\nu^c}}
\def\ecb{\overline{e^c}}
\def\lcb{\overline{\ell^c}}
\def\Ncb{\overline{N^c}}
\def\mw{m_{\rm w}}
\def\mz{m_{\rm z}}
\def\mh{m_{\rm H}}
\def\cw{c_{\rm w}}
\def\sw{s_{\rm w}}

\begin{document}

\title{\boldmath Dimension Seven Operators in Standard Model with Right handed Neutrinos }
\author{ Subhaditya Bhattacharya}
\email{subhab@iitg.ernet.in}
\affiliation{Department of Physics, Indian Institute of Technology Guwahati\\Guwahati, Assam 781039, India}
\author{Jos\'e Wudka}
\email{jose.wudka@ucr.edu}
\affiliation{Department of Physics {\it\&} Astronomy,
University of California  Riverside\\ Riverside, California 92521-0413, USA}

\begin{abstract}
In this article we consider the Standard Model extended by a number of (light) right-handed neutrinos, and assume the presence of some heavy physics that cannot be directly produced, but can be probed by its low-energy effective interactions. Within this scenario, we obtain all the gauge-invariant dimension-seven effective operators, and determine whether each of the operators can be generated at tree-level by the heavy physics, or whether it is necessarily loop generated. We then use the tree-generated operators, including those containing right-handed neutrinos, to put limits on the scale of new physics $ \Lambda $ using low-energy measurements. We also study the production of same-sign dileptons at the Large Hadron Collider (LHC) and determine the constraints on the heavy physics  that can be derived form existing data, as well as the reach in probing $ \Lambda $ expected from future runs of this collider.
\end{abstract}

%\maketitle
%\keywords{Effective Theory, Dimension seven operators, Large Hadron Collider}

%\def\skeleton{
%	- Introduction
%		? Why look for BSM physics
%		? Why use effective theories
%		? Why go up to dimension 7
%		? Why consider weakly coupled & renormalizable NP
%	- The list of operators
%		Segregate according to 
%			Type
%				PTG without n
%				PTG with n
%				LG without  n
%				LG with n 
%		Listing for each any comments, such as DB, DL and whether they vanish in the flavor diagonal case 
%	- Constraints on PTG operators
%	- LHC effects of the eeWW vertex
%	- Conclusions
%}

\maketitle
\flushbottom

\newpage

%%%%%%%%%%%  Section 1  %%%%%%%%%%%%%%%%%%%%%%%%%%%%%%%%%%%%%%

\section{\label{sec-0}Introduction}

The Standard Model (SM) is generally believed to be the low-energy limit of a more fundamental theory, however, the presence of new physics (NP) has eluded almost all experiments up to date;  the notable exceptions being the observation of neutrino masses \cite{neutrino-mass} and the very strong evidence that dark matter \cite{dark-matter} is composed of particle(s) not present in the SM. Because of this paucity of experimental guidance on even the most basic properties of NP, it is reasonable to study the effects of hypothesised heavy particles using a model-independent approach based on an effective theory. Using this approach one can derive reliable bounds (or estimates) of some of the most important parameters of physics beyond the SM (such as its scale) and to map such constraints onto specific models of NP. The procedure for constructing an \el\ is well known \cite{eff-theory1,eff-theory2, strong1}, and though some important details depend on whether the NP is assumed to be strongly \cite{strong1} or weakly \cite{eqv-th} coupled, in either case the formalism provides an efficient and consistent parameterization of all heavy-physics effects at scales below that of the NP. 

The recent observation of the Higgs boson with a mass below the electroweak scale strongly suggests that the electroweak sector of the SM is weakly coupled. This also supports the assumptions that any NP underlying the SM is also weakly coupled and, since the SM Lagrangian is renormalizable \cite{SM-renorm}, decoupling \cite{decoup}. We will adopt these assumptions in this paper, but they are  certainly not inescapable: the observed Higgs particle may not be exactly the particle predicted by the SM (e.g. the scalar sector might contain other fields) \cite{Branco:2011iw}, and it is  possible to construct models of strongly-coupled new physics that are consistent with a weakly-coupled SM \cite{Agashe:2004rs}.

The  action for the effective theory, \seff, results from integrating out all heavy modes in the full theory. By construction \seff\ will contain only SM fields, and by consistency  will respect all the SM local symmetries \cite{Veltman:1980mj}; it will also depend on the parameters of the NP and, in particular, on the typical heavy-physics scale $ \Lambda $. Expanding in powers of $\Lambda $ we can write \seff\ as the integral of a local effective Lagrangian; the decoupling assumption guarantees that terms with positive powers of $\Lambda $ are absorbed in renormalization of the low-energy theory (in this case, the SM), so that all observable NP effects are suppressed by inverse powers of $\Lambda$. Thus we can write
\beq
\leff = \lcal_{\rm SM} + \sum_{i,\,n\ge5}  \frac{c_i\up n}{\Lambda^{n-4}} \ocal_i\up n\,,
\label{eq:leff-gen}
\eeq
where the $ \ocal_i\up n $ are gauge-invariant local operators with mass dimension $n$ constructed using SM fields and their derivatives, and the $ c_i\up n $ are unknown coefficients~\footnote{If the NP Lagrangian were known, these coefficients could be calculated; absent this we take them as unknown quantities parameterizing the physics beyond the SM, and which can be experimentally measured or constrained.}. It is important to note that, though it is not indicated in the above expressions, different operators may be generated by NP of  different scales (in the above notation $ \Lambda$ can depend on the index $i$), but we will not indicate this explicitly so as not to clutter the expressions.  Although the \el\ Eq.~(\ref{eq:leff-gen}) formally contains an infinite number of coefficients, only operators of sufficienlty small dimension, corresponding to a finite number of terms, can generate effects large enough to be measured within experimental accuracy. Hence, the \el\ approach does not suffer from lack of predictability, and is useful when the nature of the NP is not known very well, as in the present situation in particle physics.

There are publications providing complete lists of operators of dimensions $ n = 5,\, 6,\ 7$  \cite{dim5,dim6-1,dim6-2,dim7-1} and partial lists of operators with $ 8\le n \le 11 $  \cite{dim8-9-10-11}; using these results very many studies have been published (see, e.g. \cite{dim7-2,eff-appl}) that obtain limits on $ \Lambda $ using a variety of processes. 

As mentioned above, one type of new physics that has been confirmed is the existence of neutrino masses, yet the mechanism responsible for them is not determined. One popular possibility is that these masses are of the Majorana type \cite{majorana}, assumed to be generated by lepton-number violating NP~\footnote{Specifically, at low energies the heavy physics is assumed to generate the dimension 5 Weinberg operator \cite{dim5} $(\phi^\dagger \ell)^2$, that produces the desired mass matrix upon \ssb.} without requiring additional light degrees of freedom. An alternative possibility is for the lepton sector to mimic the quark sector as far as mass generation is concerned, in this case it is assumed that there are 3 right-handed light neutrinos $ \nu_R$ that pair-up with their left-handed counterparts and generate Dirac masses in the usual manner.

Both cases can be studied simultaneously by including the $\nu_R$ in the set of SM fields that are to used to construct the effective Lagrangian; we will denote this model by \nsm. We emphasize that these \rh\ neutrinos are assumed to be light, the effects of heavy \rh\ neutrinos are included through the appropriate effective operators.

The list of effective operators for the \nsm\ extension of the SM are available for dimensions 5 and 6 \cite{nuSM-5, nuSM-6}. Despite the increased suppression by powers of $ \Lambda $, dimension 7 operators remain relevant because of their contributions to interesting processes, such as neutrino-less double-beta decay \cite{0nbb}, so that a complete list of dimension 7 operators for the \nsm\ will be useful in this context. The goal of this publication is to provide such a list and to analyze some of the observables  sensitive to the corresponding types of NP.

When the physics underlying the SM is weakly coupled it is useful to note that, in addition to the suppression in powers of $ \Lambda $, the effective operator coefficients will be further reduced when the corresponding operator is not generated at tree-level. It is an interesting property of renormalizable NP models that there are effective operators that are never generated at tree-level  \cite{Jose-PTG}; we will call these loop-generated (LG) operators. The remaining operators may or may not be generated at tree-level, depending on the details of the NP; we refer to these as potentially tree-generated (PTG) operators. This separation is of interest because the effects of LG operators are almost always too small to be of interest, being smaller than the 1-loop SM corrections; exceptions do occur in cases where there is no SM contribution at tree level (e.g. in Higgs production via gluon fusion  or the two-photon and $Z$-photon decay modes \cite{Higgs-LHC}). Because of this, processes to which PTG operators contribute are generally the ones most sensitive to the effects of the heavy physics \cite{Jose-PTG}. 

In practical applications of the effective theory approach, it is useful to note that the effects of some operators cannot be distinguished using only low-energy observables \cite{eqv-th}, and that this allows dropping some of the terms in Eq.~(\ref{eq:leff-gen}). Specifically, if two operators $ \ocal,\, \ocal' $ are such that the linear combination $ \ocal + r  \ocal' $ is zero on-shell, and appear in the effective Lagrangian in the combination $ c \ocal + c' \ocal'  $, then all observables will depend on $c$ and $c'$ only through the combination $ c' - r c $. In this sense, the effects of $ \ocal$ and $ \ocal' $ cannot be distinguished and either of them can be eliminated from $ \leff $ (for details see \cite{dim6-2,eqv-th})); this result is often referred to as the `equivalence theorem' \cite{eqv-th}). Once redundant operators are eliminated through this procedure the remaining ones constitute an irreducible basis. In choosing a basis  it is usually more useful to select the ones with the largest number of PTG operators (for a full discussion see \cite{Jose-basis}).

\section{\nsm\ Dimension 7 operators}
\label{sec:d7_ops}

In this section we provide a complete list of dimension 7 effective operators within the \nsm. This list was obtained in a straightforward though tedious way, beginning from a general combination of fields and ensuring Lorentz and gauge invariance; the equations of motion  were then used to eliminate redundant operators by applying the equivalence theorem \cite{eqv-th,Jose-basis}. For each of the operators listed we will also indicate whether they are LG or PTG (as noted above, this is a relevant classification for the case where the underlying physics is weakly coupled, decoupling and renromalizable); in appendix \ref{sec:lg.ptg} we present the arguments we used to obtain this classification. 

In the expressions below $ \phi $ denotes the SM scalar isodoublet, $\ell$ a \lh\ lepton isodoublet, $q$ a left-handed quark doublet, $e,\,\nu$  \rh\ charged and neutral leptons respectively (we drop the subindex $R$ to simplify the notation), and $u,\,d$  \rh\ up and down-type quarks respectively; we will for the most part suppress generation indices. We use $D$ for the covariant derivatives, and denote the $\su3_c,\, \su2_L$ and $ \ui_Y$ gauge fields by $G,\,W$ and $B$, respectively. We will also use the shorthand
\beq
N = \phi ^{\dagger}\eps\ell \,; \qquad E= \phi^\dagger\ell \,,
\eeq
motivated by the fact that in the unitary gauge $ N = (v/\sqrt{2}) \nu_L + \cdots $ and $ E = (v/\sqrt{2}) e_L + \cdots$; we also use
 \beq
 \eps=-i\tau_2 =  {\bpm 0&1 \cr -1& 0 \epm}\,,
\eeq
 where $\tau_2$ is the usual Pauli matrix and $ v \sim 246 \gev$ the \vev\ of the SM scalar doublet.

It is straightforward to show that in the \nsm\ there are no dimension 7 operators without fermions; the operators containing 2 and 4 fermions are listed below.

\subsection{Operators with 2 fermions}

These operators  are of the form~\footnote{Field strength tensors correspond to $[D,D]$ commutators contained in terms with $s=2$ in Eq.~(\ref{eq:2fops}). }
\beq
\psi^T C \Gamma \psi' \vp^r D^s\,, \quad r+s=4,~ r,s \ge0\,,
\label{eq:2fops}
\eeq
where $ \vp $ denotes $\phi $ or $ \phit=\epsilon \phi^*$, $ \psi $ a fermion in the \nsm,
\beq
\psi \in \{ q,\,u,\,d,\,\ell,\,e,\,\nu,~q^c,\,u^c,\,d^c,\,\ell^c,\,e^c,\,\nu^c \}\,,
\eeq
$C$ is Dirac charge conjugation matrix, the charge conjugate fields are defined as $\psi^c = C \bar\psi^T$, and $\Gamma=\{ 1,\gamma^\mu, \sigma^\mn\}$, where $\sigma^\mn=\frac{i}{2}[\gamma^\mu, \gamma^\nu]$. All these operators conserve baryon number but violate lepton number by two units: $|\Delta L|=2,\, \Delta B=0$ (for an interesting discussion on operators with $|\Delta (B-L)|=2$, neutrino masses and grand unification, see \cite{B-L}). 

\bit
\item{$r=4,~s=0$. 2 PTG operators:}
\beq
(\Ncb N) |\phi|^2 ,\quad \ncb\nu |\phi|^4\,.
\label{eq:r4s0}
\eeq

\item{$r=3,~s=1$. 4 PTG operators:}
\beq
(\ecb \gamma^\mu N )(\ptd \stackrel \leftrightarrow D_\mu \phi), \quad
(\ncb \gamma^\mu N )(i \phi^\dagger \stackrel \leftrightarrow D_\mu \phi), \quad
(\ncb \gamma^\mu E )(\ptd \stackrel \leftrightarrow D_\mu \phi), \quad 
(\ncb \gamma^\mu N)( \partial_\mu |\phi|^2 )\,.
\label{eq:r3s1}
\eeq
where $\phi^\dagger \stackrel \leftrightarrow D_\mu \phi =\phi^\dagger D_\mu\phi - ( D_\mu\phi)^\dagger \phi$. 
\item{$r=s=2$: 9 PTG operators.} 
\beq
\begin{array}{lll}
 ( \lcb D_\mu \ell)( \ptd D^\mu \phi), &\quad \Ncb(D_\mu\ptd  D^\mu \ell),  &\quad (\lcb D\phi) (\ell \eps D\phi), \cr
  [\Ncb \sigma^\mn (\ptd \WW_\mn \ell)], &\quad (\ncb D_\mu e)(\ptd D^\mu \phi), &\quad (\ncb \nu) |D\phi|^2, \cr
    (\ncb \sigma^\mn e)(\ptd \WW_\mn \phi) , & \quad(\Ncb\sigma^\mn N) B_\mn, &\quad  |\phi|^2 (\ncb \sigma^\mn \nu) B_\mn  ;
\end{array}
\label{eq:r2s2}
\eeq
where $\WW_\mn = \tau^I  W_\mn^I$.

\item{$r=1,~s=3$. 8 LG operators:}
\beq
\begin{array}{llll}
(\partial^\mu \ncb) \gamma^\nu N B_\mn, & \quad\ncb\gamma^\mu (\ptd D^\nu \ell) B_\mn, & \quad
(\partial^\mu \ncb) \gamma^\nu (\ptd  \WW_\mn \ell)  , &\quad \ncb\gamma^\mu (\ptd  \WW_\mn D^\nu \ell),  \cr
(\partial^\mu \ncb) \gamma^\mu N \tilde B_\mn,  & \quad\ncb\gamma^\mu (\ptd  D^\nu \ell)\tilde  B_\mn,  &\quad
(\partial^\mu \ncb) \gamma^\mu (\ptd \WW_\mn \ell) , &\quad \ncb\gamma^\mu (\ptd  \tilde\WW_\mn D^\nu \ell);
\end{array}
\eeq
where $\tilde X_\mn=\frac{1}{2}\epsilon_{\mu\nu\rho\sigma}X^{\rho\sigma}$ denote the dual tensors.

\item{$r=0,~s=4$. 6 LG operators:}
\bea
&& \ncb\nu \times \{ ( G^A_\mn)^2,\, (W^I_\mn)^2,\, (B_\mn)^2,\,( \tilde G^A_\mn G^A_\mn),\, (\tilde W^I_\mn W^I_\mn),\, (\tilde B_\mn B_\mn) \}.
\eea
\eit

\subsection{Operators with 4 fermions}

These operators are of the form $ \psi^4 D$ (operators with 4 fermions and one covariant derivative) or $ \psi^4 \varphi $ (operators with 4 fermions and one scalar); they all violate $|B-L|$ by two units with $ |\Delta B|=0,\,1$.

\bit
\item{$\psi^4 D $: 21 LG operators.}
Using Fierz rearrangements these can be cast in either of two forms:
\beq
(L_1 \sigma^\mn L_2)(L_3 \gamma_\nu \stackrel \leftrightarrow{D_\mu} R)\,, \qquad (L_1 \sigma^\mn L_2) D_\mu (L_3 \gamma_\nu R)\,;
\label{eq:4f.D}
\eeq
where $L$ and $R$ denote, respectively, left and right-handed fermion fields. The allowed field combinations are listed in table \ref{tab:4f.D}.

\begin{table}[ht]
$$
%{\tiny 
\renewcommand{\arraystretch}{1.5}
\begin{array}{|c|cccc|cc|}
%\hline
% \multicolumn{7}{c}{\ocal=(L_1 \sigma^\mn L_2)(L_3 \gamma_\nu \stackrel \leftrightarrow D_\mu R),~(L_1 \sigma^\mn L_2) D_\mu (L_3 \gamma_\nu R)}  \cr
\hline
%	& \multicolumn{4}{c|}{fields}					&	\Delta L	&	\Delta B	  \cr
	&	L_1		&	L_2		&	L_3		&	R	&	\Delta L	&	\Delta B			\cr
\hline
1	&	d^c		&	d^c		&	d^c		&	e	&	 1	&	 -1					\cr
2	&	d^c		&	\ell		&	\ell		&	u	&	 2	&	 0					\cr
3	&	d^c		&	\ell		&	d^c		&	q^c	&	 1	&	-1					\cr
&&&&&&\cr
4	&	q		&	d^c		&	\ell		&	\nu	&	 2	&	 0					\cr
5	&	q		&	\ell		&	d^c		&	\nu	&	 2	&	 0					\cr
6	&	d^c		&	\ell		&	q		&	\nu	&	 2	&	 0					\cr
7	&	\ell		&	e^c		&	\ell		&	\nu	&	 2	&	 0					\cr
8	&	q		&	u^c		&	\nu^c	&	\ell^c	&	-2	&	 0					\cr
9	&	q		&	\nu^c	&	u^c		&	\ell^c	&	-2	&	 0					\cr
10	&	u^c		&	\nu^c	&	q		&	\ell^c	&	-2	&	 0					\cr
11	&	u^c		&	\nu^c	&	e^c		&	d	&	-2	&	 0					\cr
12	&	u^c		&	e^c		&	\nu^c	&	d	&	-2	&	 0					\cr
13	&	\nu^c	&	e^c		&	u^c		&	d	&	-2	&	 0					\cr
14	&	u^c		&	d^c		&	d^c		&	\nu	&	 1	&	-1					\cr
15	&	q		&	\nu^c	&	q		&	d	&	-1	&	 1					\cr
&&&&&&\cr
16	&	q		&	\nu^c	&	\nu^c	&	q^c	&	-2	&	 0					\cr
17	&	u^c		&	\nu^c	&	\nu^c	&	u	&	-2	&	 0					\cr
18	&	d^c		&	\nu^c	&	\nu^c	&	d	&	-2	&	 0					\cr
19	&	\ell		&	\nu^c	&	\nu^c	&	\ell^c	&	-2	&	 0					\cr
20	&	\nu^c	&	e^c		&	\nu^c	&	e	&	-2	&	 0					\cr
&&&&&&\cr
21	&	\nu^c	&	\nu^c	&	\nu^c	&	\nu	&	-2	&	 0					\cr
\hline
\end{array}
%} % end of tiny
$$

\caption{ Field combinations that can contribute to the operators Eq.~(\ref{eq:4f.D}) containing 4 fermions, one derivative and no scalar fields.}
\label{tab:4f.D}
\end{table}

\item{$\psi^4 \phi $: 33 PTG operators.}
Using Fierz transformations one can readily see that these take one of the two forms:
\beq
(L_1^T C L_2)(L_3^T C L_4) \vp\,, \qquad (L_1^T C L_2)(R_1^T CR_2) \vp\,;
\label{eq:4f.phi}
\eeq
where $ \varphi = \phi,\, \phi^c$. The allowed field combinations are listed in table \ref{tab:4f.phi}.

\begin{table}[ht]
%{\tiny
$$
\renewcommand{\arraystretch}{1.5}
\begin{array}{|c|cccc|cc|c|}
\cline{1-8}
	 \multicolumn{7}{|c}{\ocal=(L_1^T C L_2)(L_3^T C L_4) \vp }		&			  \cr
\cline{1-8}
%	& \multicolumn{4}{c|}{fields}		&	\multicolumn{2}{c|}{quantum~numbers}	&&\cr
	&	L_1	&	L_2	&	L_3	&	L_4	&	\Delta L	&	\Delta B	  & \vp \cr
\cline{1-8}
1	&	\ell	&	\ell	&	\ell	&	e^c	&	 2	&	 0	&  \phi  \cr
2	&	q	&	d^c	&	\ell	&	\ell	&	 2	&	 0	&  \phi  \cr 
3^{**}	&	q	&	\ell	&	\ell	&	d^c	&	 2	&	 0	&  \phi \cr 
4	&	u^c	&	d^c	&	d^c	&	\ell	&	 1	&	-1	&  \phi  \cr 
5	&	d^c	&	d^c	&	d^c	&	\ell	&	 1	&	-1	&  \phit \cr 
6	&	u^c	&	\ell	&	d^c	&	d^c	&	 1	&	-1	&  \phi  \cr 
&&&&&&&\cr
7	&	q	&	u^c	&\nu^c	&	e^c	&	-2	&	 0	&  \phit \cr 
8	&	q	&	e^c	&\nu^c	&	u^c	&	-2	&	 0	&  \phit \cr 
9^*	&	q	&	q	&	q	&\nu^c	&	-1	&	 1	&  \phit \cr 
&&&&&&&\cr
10	&	q	&	u^c	&\nu^c	&\nu^c	&	-2	&	 0	&  \phi  \cr 
11	&	q	&	d^c	&\nu^c	&\nu^c	&	-2	&	 0	&  \phit \cr 
12	&	q	&\nu^c	&\nu^c	&	u^c	&	-2	&	 0	&  \phi  \cr 
13	&	q	&\nu^c	&\nu^c	&	d^c	&	-2	&	 0	&  \phit \cr 
14	&	\ell	&	e^c	&\nu^c	&\nu^c	&	-2	&	 0	&  \phit \cr 
15	&	\ell	&\nu^c	&\nu^c	&	e^c	&	-2	&	 0	&  \phit \cr 
&&&&&&&\cr
16	&	\ell	&\nu^c	&\nu^c	&\nu^c	&	-2	&	 0	&  \phi  \cr 
\cline{1-8}
\end{array}
\qquad \qquad
\renewcommand{\arraystretch}{1.5}
\begin{array}{|c|cccc|cc|c|}
\hline
	 \multicolumn{7}{|c}{\ocal=(L_1^T C L_2)(R^T_1 CR_2) \vp }		&				  \cr
\cline{1-8}	
%& \multicolumn{4}{c|}{fields}		&	\multicolumn{4}{c|}{quantum~numbers}	&& \cr
	&    L_1	&	L_2	&	R_1	&	R_2	&	\Delta L	&	\Delta B	& \vp \cr
\cline{1-8}
1	&	d^c	&	\ell	&	u	&	e	&	 2	&	 0	 & \phi  \cr
2	&	\ell	&	\ell	&	q^c	&	u	&	 2	&	 0				 & \phi  \cr
3^*	&	q	&	q	&	d	&	\ell^c	&	-1	&	 1				 & \phit \cr
4	&	q	&	e^c	&	d	&	d	&	-1	&	 1				 & \phit \cr
&&&&&&&\cr
5	&	q	&	d^c	&	\nu	&	e	&	 2	&	 0			&	  \phi  \cr
6	&	u^c	&	\ell	&	u	&	\nu	&	 2	&	 0			&	  \phi  \cr
7	&	d^c	&	\ell	&	u	&	\nu	&	 2	&	 0			&	  \phit \cr
8	&	d^c	&	\ell	&	d	&	\nu	&	 2	&	 0			&	  \phi  \cr
9	&	\ell	&	e^c	&	\nu	&	e	&	 2	&	 0			&	  \phi  \cr
%10	&	d^c	&	\ell	&	u	&	\nu	&	 2	&	 0			&	 \phit \cr
%10	&	\ell	&	e^c	&	\nu	&	e	&	 2	&	 0			&	  \phi \cr
10^{**}	&	q	&	\ell	&	q^c	&	\nu	&	 2	&	 0			&	 \phi \cr
11	&	\ell	&	\ell	&	\ell^c	&	\nu	&	 2	&	 0			&	  \phi  \cr
12	&	q	&\nu^c	&	u	&	d	&	-1	&	 1			&	  \phit \cr
13	&	q	&\nu^c	&	d	&	d	&	-1	&	 1			&	  \phi  \cr
&&&&&&&\cr
14	&	q	&	u^c	&	\nu	&	\nu	&	 2	&	 0			&	  \phi  \cr
15	&	q	&	d^c	&	\nu	&	\nu	&	 2	&	 0			&	  \phit \cr
16	&	\ell	&	e^c	&	\nu	&	\nu	&	 2	&	 0			&	  \phit \cr
&&&&&&&\cr
17	&	\ell	&\nu^c	&	\nu	&	\nu	&	 2	&	 0			&	  \phi  \cr
\cline{1-8}
\end{array}
$$
%} % end of \tiny
\caption{Possible field combinations appearing in the four fermion operators containing one scalar and no derivatives Eq.~(\ref{eq:4f.phi}). The entries with one (two) asterisks have 2 (3) possible $\su2$ contractions (assuming only family-diagonal couplings, see text).}
\label{tab:4f.phi}
\end{table}

\eit

The list of operators provided here do not include family labels to avoid notational clutter. In certain cases, however, the operators vanish when some of the fields are in the same family. For example, it is easy to see that $ \ell^ c C \ell = 0 $ when both lepton isodoublets are in the same family, so any operators with this factor will not have family-diagonal contributions and should in principle be written as $ \ell_i^ c C \ell_j$ where $i,j= 1,2,3~ (i \ne j)$ denote family indices. The contraction of the $ \su3 $ colour indices is unambiguous in the above operators since it can take only two forms: either $\{Q^c_r \,Q'_s  \}\delta_{rs} $ or $\{ Q_r^c \,Q_s'{}^c \,Q_u''{}^c \} \epsilon_{rsu}$, with $Q,\,Q', Q''$ denoting generic quark fields and $r,s,u$  colour indices.

\section{PTG operators that do not contain \rh\ neutrino fields}
\label{sec: SMops}

We will consider separately operators that contain \rh\ neutrinos in section \ref{sec:rhn} below; here we will discuss the leading effects of dimension 7 operators containing only SM fields and which can be generated at tree level. There are 20 such operators:
\beq
{\small
\begin{array}{llll}
\ocal_1 = (\lcb \eps D^\mu \phi) (\ell \eps D_\mu \phi),\quad & % weak bounds e.g. Z --> W e nu
\ocal_2 = (\ecb \gamma^\mu N )(\phi \epsilon  D_\mu \phi),\quad &   % W decays 400 GeV
\ocal_3 = ( \lcb \eps D_\mu \ell)( \phi \eps D^\mu \phi),\quad &  % W decays 35 geV
\ocal_4 = \Ncb (D_\mu \phi \eps D^\mu \ell),\cr % Z decays 92 GeV
\ocal_5 = (\Ncb \ell )\eps(	\bar e \ell	),\quad &  % 182 geV from ee--> nu nu assuming th esame error as in Z--> invisible decays
\ocal_6 = (\Ncb N) |\phi|^2,\quad &  % nu masses 770 TeV
\ocal_7 = [\Ncb \sigma^\mn (\phi\eps\WW_\mn\ell)],\quad &   % plasmon decay 33 TeV
\ocal_8 = (\Ncb \sigma^\mn N) B_\mn ,\cr% plasmon decay 47 TeV
\ocal_9 = (\bar d q	)\eps(	\Ncb \ell	),\quad &  % pi+ decay 2.1 TeV
\ocal_{10} = [(\overline{q^c} \phi)	\eps \ell )( \bar d \ell ),\quad &  % pi+ decay 2.1 TeV
\ocal_{11} = (\Ncb q )\eps(	\bar d \ell	),\quad &   % pi+ decay 2.1 TeV
\ocal_{12} = (\lcb	\eps q )( \bar d N ),\cr   % pi+ decay 2.1 TeV
\ocal_{13} = (\bar d N )(	u^T C e ),\quad &  % pi+ decay 2.1 TeV
\ocal_{14} = (\Ncb \ell	)( \bar q u ),\quad &  % pi+ decay 2.1 TeV
\ocal_{15} = (\bar u d^c )(	\bar d N ),\quad & % N --> pi nu 1.9 x 10^8 TeV
\ocal_{16} = [\overline{q^c}  (\phi^\dagger q) ]\eps( \bar\ell d ),\cr % N --> pi nu 1.9 x 10^8 TeV
\ocal_{17} = (\overline{q^c} \eps q	)( \bar N d	),\quad &  % N --> pi nu 1.9 x 10^8 TeV
\ocal_{18} = (\bar d d^c )(	\bar d E ),\quad & % n --> e K 9.6 x 10^7 TeV
\ocal_{19} = (\bar e  \phi^\dagger q)(	\overline{d^c}	d	) ,\quad  &  % n --> e K   9.6 x 10^7 TeV
\ocal_{20} = (\bar u N	)( \bar d d^c	)  .% n --> nu K^0_S 1.5 x 10^8 TeV
\end{array}
\label{eq:PTG-nonu}
} % end of \small
\eeq
This list coincides with the one previously presented in the literature \cite{dim7-1}.

Using the results of appendix \ref{sec:lg.ptg}  it is a straightforward exercise to determine the types of new physics that can generate those operators at tree level.

\subsection{Constraints on PTG operators without right-handed neutrinos}
\label{sec:decays}

There are a variety of existing data that can be used to constrain the scale of new physics responsible for the operators being considered here. In this section we provide limits for the PTG operators listed in Eq.~(\ref{eq:PTG-nonu}); in obtaining the numbers below we assumed no deviations from the SM and took $3 \sigma $ intervals. Though there are processes that can receive contributions form more than one operator we will provide limits for the most conservative case where there are no interference effects or cancellations (if this is relaxed the restrictions can be much weaker); we also assumed that the gauge bosons are universally coupled (so they always appear multiplied by the corresponding gauge coupling). All limits below are on $ \tilde\Lambda =  \Lambda /f^{1/3} $ that translate into limits on the new physics scale with the additional naturality assumption $ f \sim 1 $ (for weakly-coupled heavy physics). We also provide separately the limits derived from neutrinoless double beta decay experiments (for a recent review see \cite{Vergados:2012xy}) to illustrate the importance of this high precision measurement.

\bit
\item There are no published limits on $ \tilde\Lambda$ for $ \ocal_1$ from collider or gauge-boson decay data. This operator, however, contributes to neutrinoless double-beta decay \cite{delAguila:2012nu} that gives the limit $ \tilde \Lambda > 7.5~\tev $.

\item The strictest limits  on $ \tilde\Lambda$ for $ \ocal_{2,3,4,5}$ from  gauge boson decays \cite{PDG} are: $\ocal_2$:~400~\gev~$(W \to \ell \nu)$; $\ocal_3$:~35 ~\gev~$(W \to \ell \nu)$; $\ocal_4$:~92~\gev~$( Z \to \nu \nu )$; $\ocal_5$:~182~\gev~$(ee \to \nu\nu)$, assuming the same error as in the invisible decay width of the $Z$. The best limits on $\ocal_{2,3,4}$ are derived form their contribution to neutrinoless double-beta decay: $\ocal_2$:~106~\tev;
$\ocal_{3,4}$:~7.5~\tev.

\item The strictest limits  on $ \tilde\Lambda$ for $ \ocal_6$ from neutrino mass constraints is $ 770~\tev $ \cite{PDG}. The limit from neutrinoless double-beta decay is significantly stronger: $\tilde \Lambda > 2,200 ~\tev$.

\item The strictest limits  on $ \tilde\Lambda$ for $ \ocal_{7,8}$ are obtained from limits on red-giant cooling \cite{plasmon} generated by  plasmon decay $ \gamma \to \nu\nu $, which is mediated by these operators. Using the results in \cite{nuSM-5} we find
$\ocal_7: 33 ~\tev$;
$\ocal_8 : 47 ~\tev$. The neutrinoless double-beta decay on $ \ocal_7 $ is weaker: $\tilde\Lambda > 7.5~\tev $.

\item The limit on $ \tilde\Lambda$ for $ \ocal_{9,10,11,12,13,14}$ obtained from $ \pi \to e \nu $ decay is $ 2.1~\tev $ \cite{PDG}. The limit derived from neutrinoless double-beta decay is stronger: $\tilde\Lambda > 137~\tev $.

\item The strictest constraint on $ \tilde\Lambda$ for the baryon-number violating operators $ \ocal_{15,16,17}$  is obtained from the limit on $ n \to \pi \nu $ decay and give $ 1.9  \times 10^8 ~\tev $  \cite{PDG}.

\item The strictest limits  on $ \tilde\Lambda$ for the baryon-number violating operators $ \ocal_{18,19,20}$ are obtained from neutron decay:
$\ocal_{18,19} : 9.6 \times 10^7 ~\tev \,  (  n \to e K )$;
$\ocal_{20} : 1.5 \times 10^8 ~\tev \, ( n \to \nu K^0_S)$  \cite{PDG}.

\eit

It is worth noting that any such new physics that generates the operators $ \ocal_{1,3,4}$ at tree level necessarily generates the dimension 5 operator $ \ocal\up5 = \overline{N^c}N$ also at tree level \cite{delAguila:2012nu}, and the limits on $ \Lambda $ derived from the latter are much stronger: $ 10^{11} ~\tev$ (assuming $O(1)$ couplings). 

As can be seen from the above results the PTG operators in Eq.~(\ref{eq:PTG-nonu}) without right-handed neutrinos but containing quarks are highly constrained from various precision measurements and astrophysical observations. These results rely heavily on the assumption that there are no interference effects among the various operator contributions, when these are present the above limits can be significantly degraded (such cancellations may result form some unknown symmetry and are not necessarily from fine-tuning. Because of this, probing the individual operator effects (see the following section for an example) is of importance in mapping potential NP contributions, even though this usually provides weaker limits.

\subsection{Neutrino Majorana masses}

The PTG operators $ \ocal_{1-14}$ in Eq.~(\ref{eq:PTG-nonu}) generate neutrino Majorana masses through radiative corrections \cite{delAguila:2012nu}, which have the generic form $ m_{\nu - {\rm Maj}} \sim v^2/( 16\pi^2 \Lambda) $, multiplied in some cases by a SM Yukawa coupling. A detailed phenomenological investigation of the consequences of these effects is best done within the context of specific models, since correlations between effective operator contributions can be important (see {\it e.g.} \cite{delAguila:2011gr}); for example, the operator coefficients for $\ocal_{1,3,4,6-8}$ may contain Yukawa couplings that mix heavy and light fermions, whose impact cannot be gauged within  this effective approach. We will then restrict ourselves to displaying the generic expressions obtained using straightforward estimates:
\beq
\begin{array}{l|l}
{\rm Operator(s)} & m_{\nu-{\rm Maj.}} ~{\rm estimate} \cr
\hline
\ocal_{1,3,4,6-8} & v^2/(16\pi^2 \Lambda) \cr
\ocal_{2,5} & v m_e/(16\pi^2 \Lambda) \cr
\ocal_{9-12} & v m_d/(16\pi^2 \Lambda) \cr
\ocal_{14} & v m_u/(16\pi^2 \Lambda)
\end{array}
\label{eq:mnu-maj}
\eeq
where $ m_{e,d,u}$ denote the masses of light charged leptons, down and up quarks, respectively; $ \ocal_{13} $ contributes only at two loops, and  $ \ocal_{15-20}$, being baryon-number violating, generate contributions only through graphs quadratic in the effective operators.

A measure of care should be exerted in obtaining the  contributions from $\ocal_{1,3,4,6-8}$: for example, after \ssb\ and in unitary gauge,
\beq
\ocal_1 \supset \half \left[ \left( \partial H \right)^2 - m_{\rm Z}^2 Z^2 \right] \left( \overline{ \nu_L^c} \, \nu_L \right)\,,
\eeq
which generates two contributions to $ m_{\nu - {\rm Maj}}$, one from a Higgs ($H$) loop and another a $Z$. Each loop gives $ \Lambda/(16\pi^2) $ to leading order, but they cancel, leaving only the sub leading contribution listed above. This cancellation is not accidental: an examination of the operators shows that, absent the spontaneous breaking of the SM gauge symmetry, they do not generate one-loop contributions to $ m_{\nu - {\rm Maj}}$.

For $ m_{\nu - {\rm Maj}} \sim 0.1~{\rm eV} $, we obtain from Eq.~(\ref{eq:mnu-maj}) $\Lambda \sim 8 \times 10^3 - 4 \times 10^9 ~\tev$ depending on the operator used, and assuming all operator coefficients are $O(1)$.

\section{Example of an LHC effect}

Consider the PTG operators $\ocal_1=(\lcb \eps D^\mu \phi) (\ell \eps D_\mu \phi) $ and $\ocal_3=( \lcb \eps  D_\mu \ell)( \phi \eps D^\mu \phi)$ from Eq.~(\ref{eq:PTG-nonu}). It is easy to see that, in unitary gauge they both contain the same lepton-number violating vertex involving two $W$ gauge bosons and two left-handed charged leptons:
\beq
( \lcb \eps  D_\mu \ell)( \phi \eps D^\mu \phi) ,~ (\lcb \eps D^\mu \phi) (\ell \eps D_\mu \phi) \supset
m_{\rm W}^2  \left(1 + \frac hv \right)^2  {W^+}^2 (e_L^T Ce_L) 
\label{op:ug}
\eeq
(no other operator in Eq.~(\ref{eq:PTG-nonu}) contains this vertex). Despite their both containing the  vertex Eq.~(\ref{op:ug}) $\ocal_{1,3}$ are generated at tree-level by different types of heavy physics (see appendix \ref{sec:lg.ptg}); this will be of use in understanding the types fo NP that can be probed through Eq.~(\ref{op:ug}); see sect. \ref{sec:appl}. 

Below we will consider the effects of the \rhs in Eq.~(\ref{op:ug}) in the production of same-sign dileptons at the LHC, and determine the constraints on the scale of new physics $ \Lambda $ that can be derived from existing data. It should be noted that $ \ocal_{1,3}$ also contain other lepton-number violating vertices in addition to the one in Eq.~(\ref{op:ug}): $ \ocal_1 \supset W\nu e,\,WZ\nu e,\, WA\nu e,\, WW\nu\nu$ and $ \ocal_3 \supset ZZ\nu\nu,\, WZ\nu e$, plus others involving the Higgs field; these will contribute to a variety of other reactions from which independent constraints on $ \Lambda $ can  be derived. Note however that such vertices involve one or more neutrinos and/or Higgs fields, and because of this the corresponding constraints will be weaker. It is for this reason that we concentrate on the term containing two charged leptons; the constraints on $ \Lambda $ using this term of course apply to all the vertices contained in $ \ocal_{1,3} $.

In the following we will define
\beq
\ocal_{\ell\ell} = f_1 \ocal_1 + f_3 \ocal_3\,,
\label{eq:oll}
\eeq
and consider separately the same-sign dilepton signal associated with two jets and the hadronically-quiet trilepton events at the Large Hadron Collider (LHC) generated by this operator. 

\subsection{Same Sign dilepton Signal at the LHC} 

\begin{figure}[thb]
$$
\begin{array}{ccc}
\vspace{-.1in}
{\includegraphics[height=3cm]{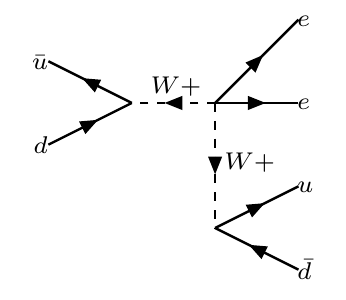}}&
{\includegraphics[height=3cm]{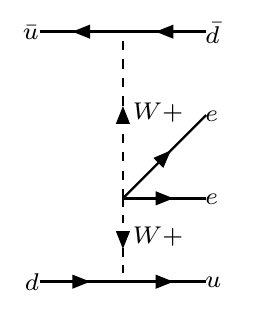}}&
{\includegraphics[height=3cm]{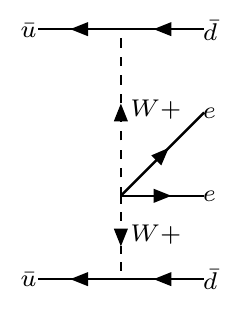}} \cr
(a) & (b) & (c) \cr
&&\cr
\vspace{-.1in}
{\includegraphics[height=3cm]{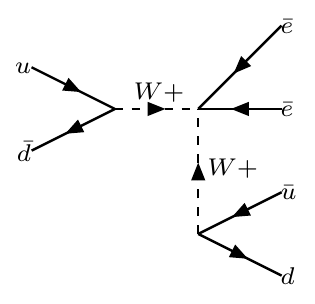}}&
{\includegraphics[height=3cm]{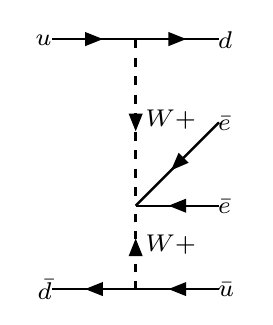}}&
{\includegraphics[height=3cm]{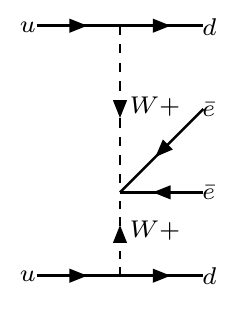}}\cr
(d) & (e) & (f)
\end{array}
$$
\caption{Leading Feynman diagrams that contribute to $pp \to \ell\ell jj$ at the LHC; the $eeWW$ vertex Eq.~(\ref{op:ug}) is generated by  $\ocal_{\ell\ell}$ in Eq.~(\ref{eq:oll}).}
\label{fig:fd12}
\end{figure}

$\ocal_{\ell\ell}$ in Eq.~(\ref{eq:oll}) will produce a dilepton signal $pp\rightarrow \ell \ell j j$ at LHC; where $\ell=e, \mu$ have the {\em same} sign, and the $j$ denote light-quark jets (after tagging efficiencies are included, the number of $ \tau $-lepton events is too small to be of interest). In Fig.~(\ref{fig:fd12}) we show the dominant Feynman diagrams that contribute to the  $\ell\ell jj$ final state generated by this operator; these can be separated into $s$-channel reactions and $t$-channel reactions; we will see that the latter dominate over the former. In the $s$-channel contributions (diagrams $a,\,d$ in the figure) one $W$ in the vertex Eq.~(\ref{op:ug}) couples to the quarks in the colliding protons, while the other couples to the light jets in the final state; in the $t$-channel processes (diagrams $b,\,c,\,e,\,f$) each $W$ couples to an incoming and an outgoing quark. 

As $\ocal_{\ell\ell}$ is of dimension 7, its coefficient contains a $\Lambda^3$ suppression factor that will prevent  probing new physics above the  TeV region, as we will shortly demonstrate. Even with this limitation we will argue that the constraints obtained are of interest. In the following we will assume the effective operator coefficients are $O(1)$, if this is not the case the limits obtained apply to the scale $ \tilde\Lambda$ introduced in section \ref{sec:decays}.

\begin{table}[ht]
%\begin{minipage}{\textwidth}   
{\small
\begin{center}
\begin{tabular}{|c|c|c|c|}
%\hline
\multicolumn{4}{c}{$E_{CM} = 14~\tev$}\cr
\hline
 Process & leading subprocesses & $\sigma$ (pb) & $\sigma$ after $C_1$ (pb)\\
\hline
\multirow{4}{*}{$pp \to e e q q$}	& $dd \to eeuu$			& 0.142	& 0.118	\\
						& $\bar{u}d \to ee\bar{d}u$	& 0.039	& 0.038	\\
						& $sd \to eeuc$				& 0.024	& 0.020	\\
						& $dd \to eeuc$				& 0.015	& 0.014	\\
\cline{3-4}
						& Total\footnote{Totals refer to the sum of all contributions, not only the leading ones.}					& 0.253	& 0.215	\\
\hline
\multirow{5}{*}{$pp\to \bar{e}\bar{e}qq$}	& $uu \to \bar{e}\bar{e}dd$			& 0.942	& 0.784	\\
								& $u\bar{d}\to \bar{e}\bar{e}d\bar{u}$		& 0.121	& 0.100	\\
								&  $uu\to \bar{e}\bar{e}ds$ 			& 0.097 	& 0.082	\\
								&  $u\bar{s}\to \bar{e}\bar{e}d\bar{c}$ 	& 0.064 	& 0.052 	\\
\cline{3-4}
								& Total 							& 1.300 	& 1.080	\\
\hline
\hline
$pp\to \ell \ell qq $ & sum of all contributions for $\ell=e,\mu$ & 3.106 & 2.590 \\
\hline
\end {tabular}
\end{center}
\vspace{.1 cm}
\begin{center}
\begin{tabular}{|c|c|c|c|}
%\hline
\multicolumn{4}{c}{$E_{CM} = 8~\tev$}\cr
\hline
Process & Subprocesses & $\sigma$ (pb) & $\sigma$ after $C_1$ (pb)\\
\hline
\multirow{4}{*}{$pp \to e e q q$}		& $dd \to eeuu$			& 0.016	& 0.012	\\
							& $\bar{u}d \to ee\bar{d}u$	& 0.004	& 0.003	\\
							& $ds \to eeuc$ 			& 0.002	& 0.002	\\
							& $dd \to eeuc$				& 0.002	& 0.001	\\
\cline{3-4}
							& Total					& 0.028	& 0.021	\\
\hline
\multirow{5}{*}{$pp\to \bar{e}\bar{e}qq$}	& $uu \to \bar{e}\bar{e}dd$			& 0.102	& 0.084	\\
								& $u\bar{d} \to \bar{e}\bar{e}d\bar{u}$	& 0.013	& 0.010 	\\
								& $uu \to \bar{e}\bar{e}ds$ 			& 0.010	& 0.009	\\
								& $u\bar{s} \to \bar{e}\bar{e}d\bar{c}$ 	& 0.006	& 0.005	\\
\cline{3-4}
								& Total 							& 0.140	& 0.115	\\
\hline
\hline
$pp\to \ell \ell qq $ & sum of all contributions for $\ell=e,\mu$  & 0.336 & 0.273 \\
\hline
%\hline
\end {tabular}
\end{center}
\caption {Total cross-section for $\ell\ell jj$ production, and of the leading contributing subprocesses at the LHC with $E_{\rm CM}=14$ and $8~\tev$ generated by the operator $ \ocal_{\ell\ell}$. The last column indicates the effects of the cuts in Eq.~(\ref{eq:c1}). We use $\Lambda = 100 ~\gev$ (see text).}
\label{table:eeqq1}
} % end of \small font
%\end{minipage}
\end{table}

We note here that such a lepton-number violating signal ($eeu\bar d, ee \bar d \bar d, \bar e \bar e \bar u d, \bar e \bar e dd $) cannot be  exclusively produced by SM, which conserves lepton number. Since the amplitude contains $1/\Lambda^3$ from $\ocal_{\ell\ell}$, the signal cross section will be proportional to $ 1/\Lambda^6$, so that 
\beq
 \sigma\up{\rm signal}_{\ell\ell jj}(\Lambda') = \left( \frac\Lambda{\Lambda'} \right)^6 \sigma\up{\rm signal}_{\ell\ell jj}(\Lambda)\,.
 \label{eq:scale}
 \eeq
 
 Thus we can compute the cross section at a convenient value of $ \Lambda$ and use this scaling property to obtain $ \sigma_{\ell\ell jj} $ for any other scale. In the following we evaluate first  $ \sigma_{\ell\ell jj} $ for $ \Lambda = 100 ~\gev $ at the LHC for $14 ~\tev$ and $8 ~\tev $  CM energies (see table \ref{table:eeqq1}). We obtained these results using the {\tt Calchep 3.6.14} \cite{CalcHEP} event generator to calculate the hard cross-sections, and we chose the CTEQ6L parton distribution function \cite{CTEQ} with the invariant mass of the two incoming quarks as the renormalisation and factorization scales. There is a variation of up to $\sim15 \%$ in the cross section when the parton distribution function, and  renormalisation and factorization scales are varied, which can presumably be addressed by a next-to-leading order calculation; this effort, however, lies beyond the scope of this investigation. We also imposed the following basic cuts:
\beq
C1: ~{p_T}_{(\ell,j)}>15\gev; \quad |\eta_{\ell}|<2.5\,,
\label{eq:c1}
\eeq
where $ {p_T}_{(\ell,j)}$ denotes the lepton and jet transverse momenta, and $\eta_{\ell}$ the lepton pseudo-rapidity. It is worth noting that there will be a  difference in the production cross-sections  for positively and negatively charged same-sign dileptons, which is due to the difference in the $u$ and $d$ parton distributions in the proton; there would be no such difference in a  $p\bar{p}$ machine.

\begin{figure}[thb]
\centering
\centerline{\includegraphics[height=5cm]{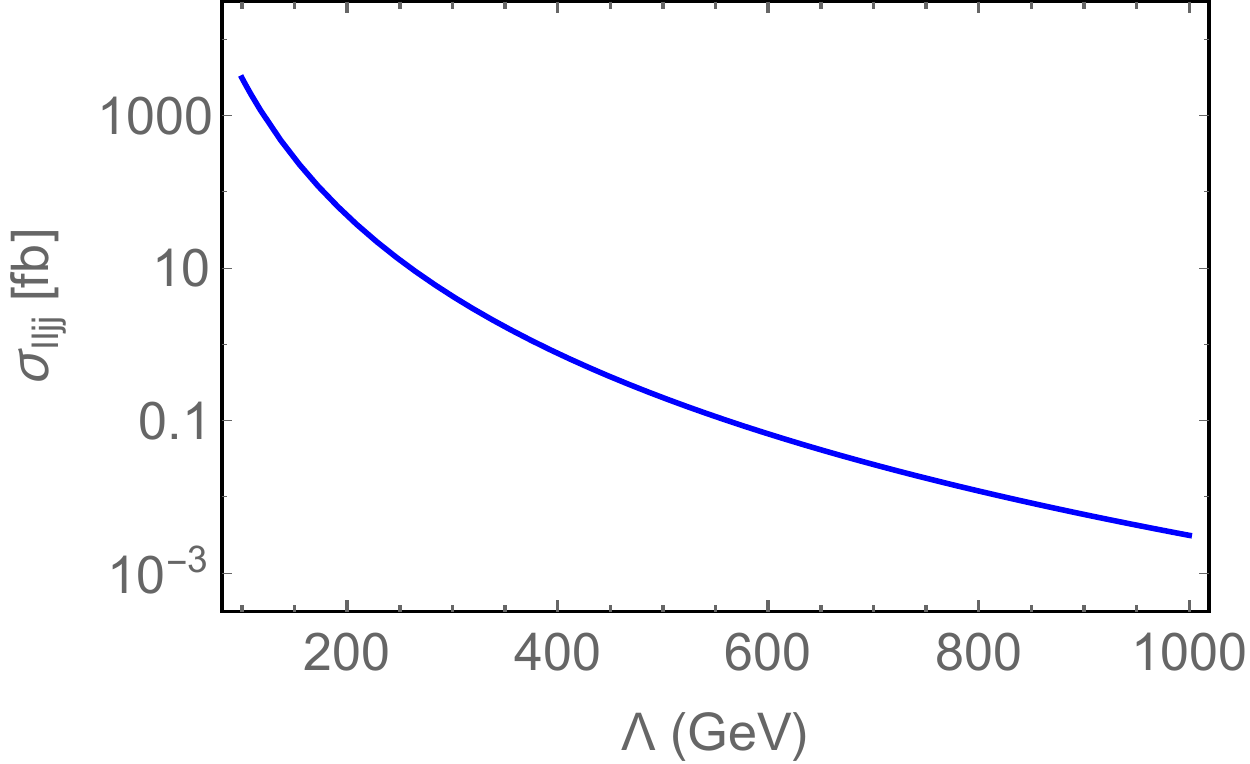}
\hskip 20pt \includegraphics[height=5cm]{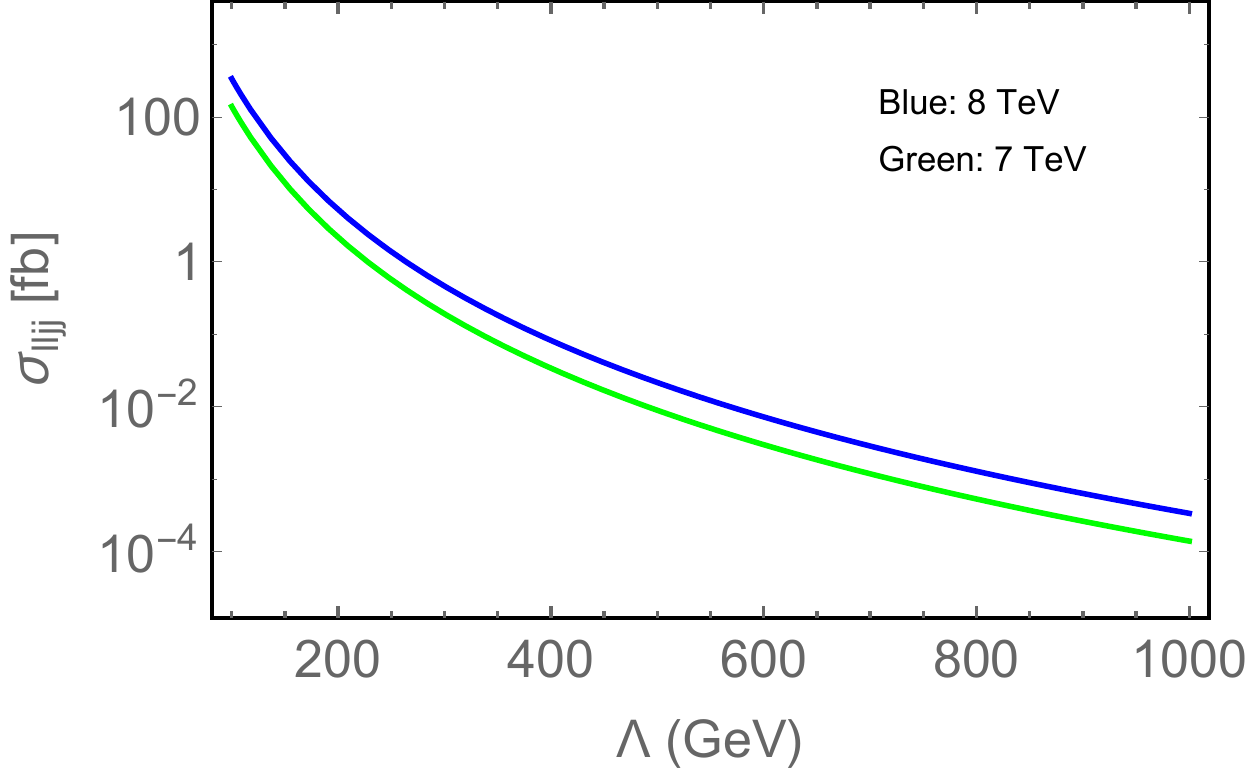}}
\caption {Production cross-section for two same-sign leptons and two jets ($\ell\ell jj$) as a function of $\Lambda $ (GeV) at LHC with $E_{\rm CM}= 14$ TeV (left) and  $7,\, 8$ TeV (right) in green and blue respectively. }
\label{fig:ssdnp}
\end{figure}

All calculations were made in the unitary gauge (again we emphasize that the choice of $\Lambda=100~\gev $ is made for calculational ease, and not with the assumption that there is new physics lurking at $0.1~\tev$); the $1/\Lambda^6$ behaviour of  $ \sigma_{\ell\ell jj} $ as $ \Lambda $ changes is presented in Fig.~(\ref{fig:ssdnp}). For example, we see that for $\Lambda=500~\gev$, $ \sigma_{\ell\ell jj} =0.165$~fb at 14 TeV CM energy (after the cuts C1 in Eq.~(\ref{eq:c1}) are imposed). This corresponds to $16$ events for an integrated luminosity of $100~ $fb$^{-1}$ (consistent with the expectations for `run 2' at the LHC \cite{ATLAS1,CMS1}); this would increase to $497$ events for the proposed high-luminosity upgrade \cite{HL-LHC} with a projected integrated luminosity of $ 3000 ~$fb$^{-1}$. For 7 and 8 TeV CM energies and $ \Lambda = 500 ~\gev $ the cross section drops to $0.011$ fb and $0.017$ fb respectively and has no observable effects.

The discovery limit, however, depends on the SM background estimate for $\ell\ell jj$ signal, which we discuss below. We will use these results to obtain the current limits on $ \Lambda $ derived form the $8 ~\tev$ LHC data, and to derive the expected sensitivity that will be reached when the CM energy is increased to $14 ~\tev$.

\begin{figure}[thb]
\centering
\centerline{\includegraphics[height=5.6cm]{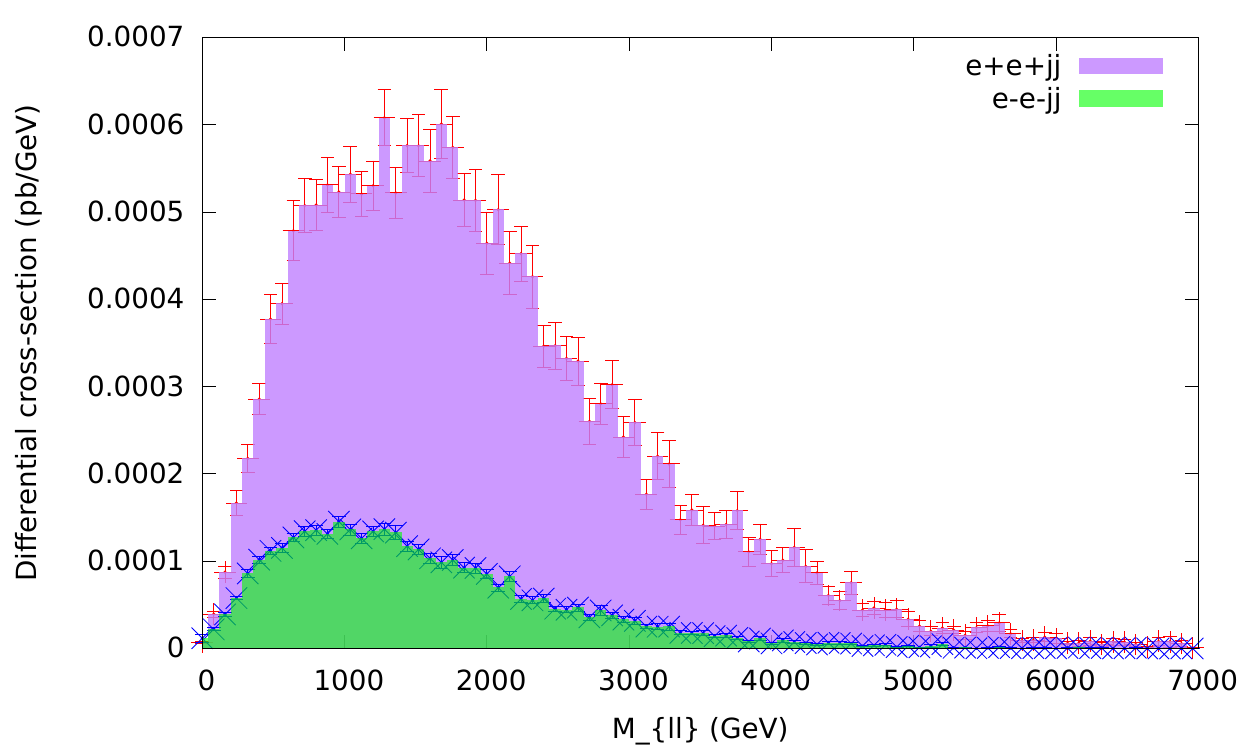}
\hskip 10pt \includegraphics[height=5.6cm]{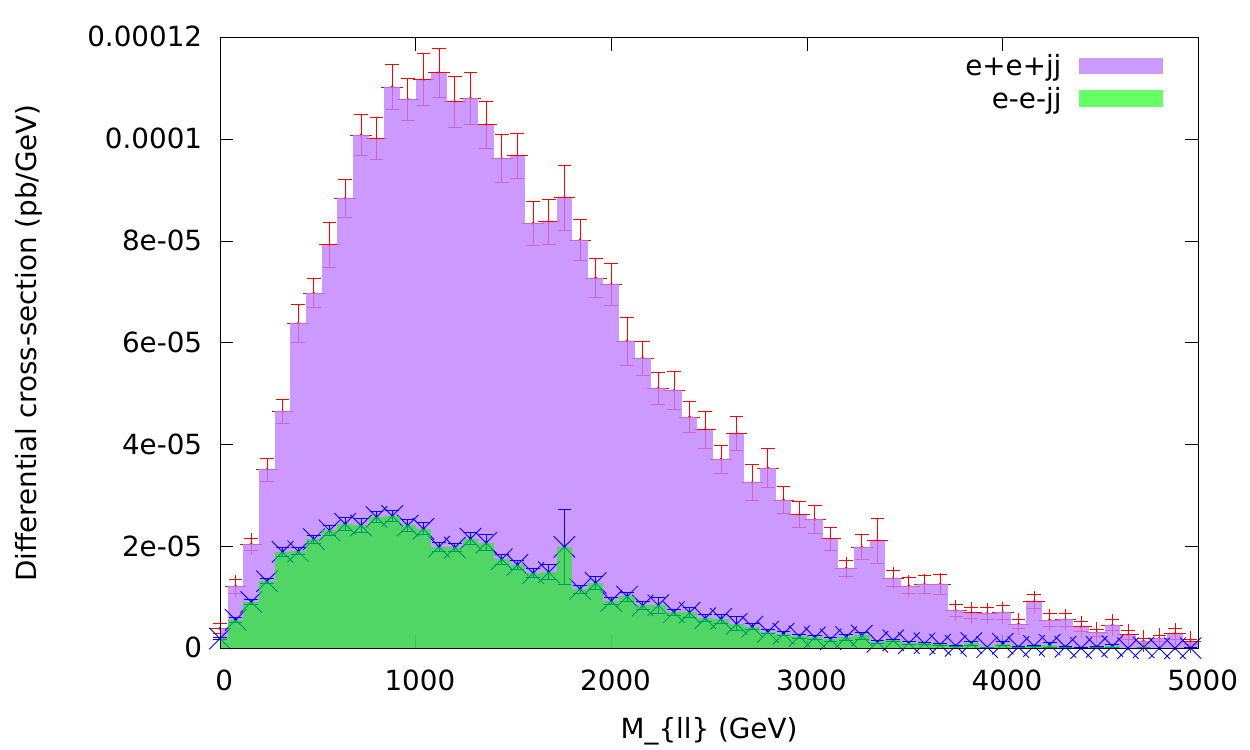}}
\vskip 10pt \centerline{\includegraphics[height=5.6cm]{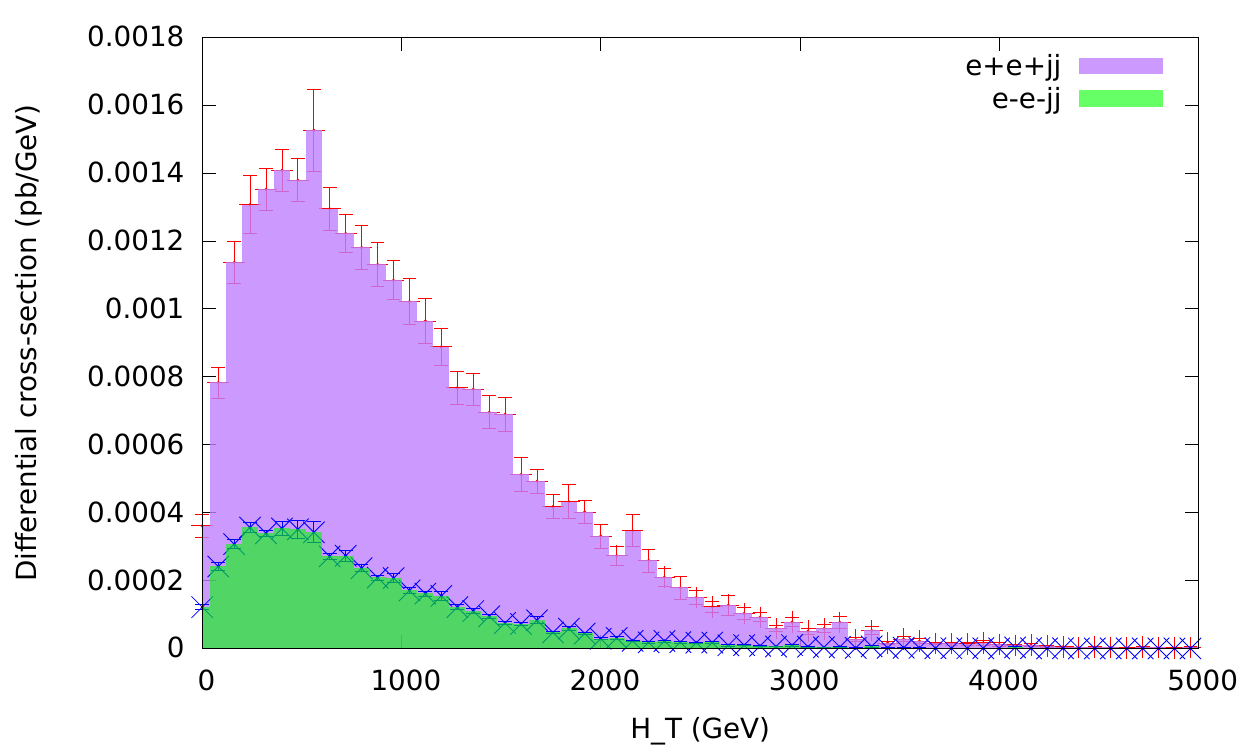}
\hskip 10pt \includegraphics[height=5.6cm]{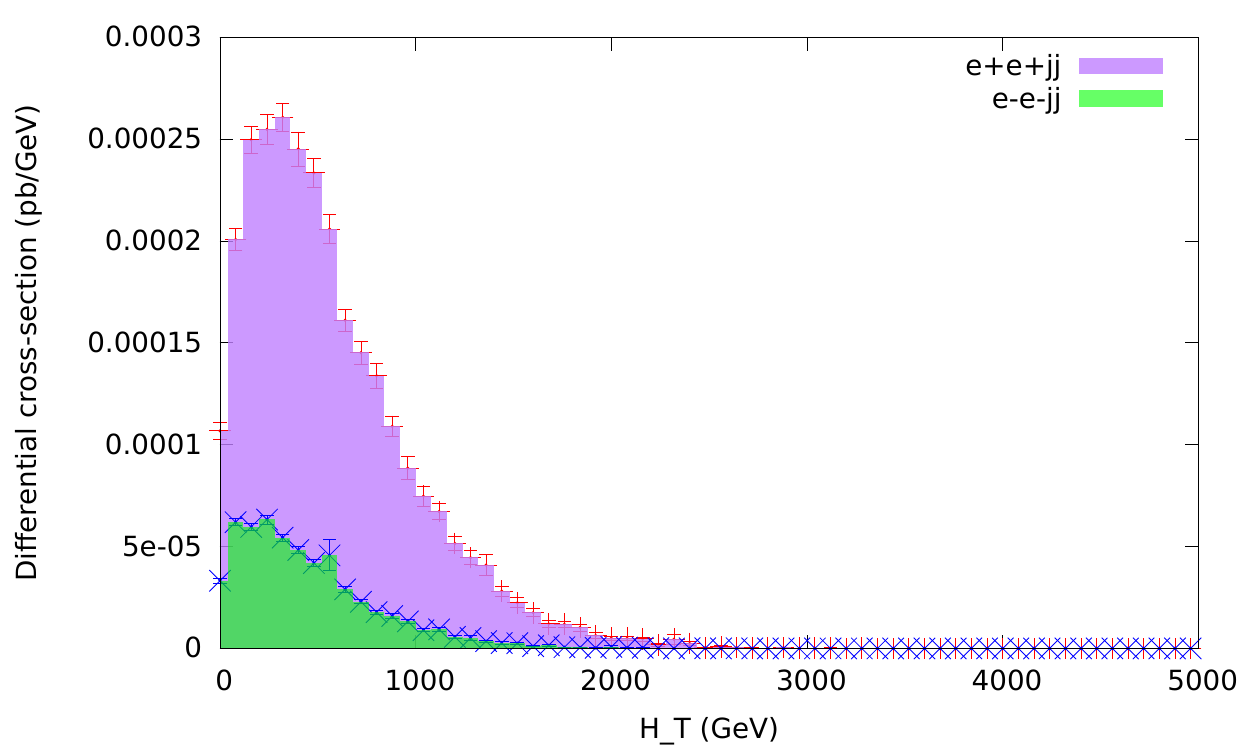}}
\caption{Distributions of the invariant mass of leptons $M_{\ell\ell}$ (top)  and transverse mass $H_T$ (bottom), for the  $\ell\ell jj$ signal events at LHC with $E_{cm}= 14$ TeV  (left) and  $E_{cm}= 8$ TeV (right).  We took $\Lambda= 100~ \gev$ (see text and Eq.(\ref{eq:scale})). Subprocesses $\bar{e} \bar{e}jj$ (red dots and lilac histograms) and $eejj$ (blue crosses and green histograms) are shown separately.}
\label{fig:sig-inv-mass}
\end{figure}

\subsubsection{SM background for same-sign dilepton signal at the LHC}

The most significant background contribution to our process is generated by SM $t\bar{t}$ production, with marginal contributions from  $t\bar{t}W$, $t\bar{t}Z$ and diboson production. 
For  the $t\bar{t}$ background calculation we take $m_{\rm top}= 173.34 ~\gev$  \cite{topmass} and use the {\tt Pythia 6.4} \cite{Pythia} event generator at tree level, which we multiply the by the appropriate $K-$factor~\footnote{The $K$ factor is used to correct the tree-level cross sections used by {\tt Pythia} so they match the NLO+NLL predictions or the experimental data; these (mostly QCD) corrections can be significant, as in the case in $ t \bar t $ production of interest here.} to obtain the NLO+NLL cross-section at the LHC \cite{ttbar-7,ttbar-8,ttbar-14}. This $K$ factor is very significant, for example, the {\tt Pythia} $t\bar{t}$ production cross-section  is $386.8$~pb  at $14$ TeV, while, the NNLO prediction is between $805$ and $898$ pb (incorporating the jet energy scale and parton distribution function  uncertainties); we use the average value of $851.5$ pb that gives  $K=2.2$ factor. For $E_{CM}= 8$ TeV, the {\tt Pythia} prediction is $94$ pb, while the measured value \cite{ttbar-8} in the dilepton and lepton+jet channels is $161.9$ pb (there are variations in this number depending not he channel used) so that  we use $K=1.7$ in this case. 

In our analysis we try to mimic the experimental reconstruction for leptons and jets within {\tt Pythia} by imposing the following requirements:
\bit
\item {\em Leptons} ($\ell$) are identified as electrons and muons with transverse momentum $p_T>$ 10 GeV and rapidity $|\eta|<$2.5. Two leptons with pseudo-rapidities $\eta$ and $ \eta + \Delta \eta $, and azimuthal angles $ \phi$ and $ \phi + \Delta \phi $ will be considered isolated if $ \Delta R = \sqrt{(\Delta \eta)^2 + (\Delta\phi)^2} \ge 0.2 $. A lepton and a jet will be isolated if $\Delta R \ge 0.4 $ and if the $ \Delta R \le0.2 $ cone contains less than $10$ GeV of transverse energy from low-$E_T$ hadron activity

\item {\em Jets} ($j$) are formed with all the final state particles after removing the isolated leptons from the list with {\tt PYCELL}, a built-in cluster routine within {\tt Pythia}. The detector is assumed to span  the pseudorapidity range $|\eta| \le5$ and to be segmented in 100-$\eta$ and 64-$ \phi $  bins. The minimum transverse energy $E_T$ of each cell is taken as $0.5$ GeV, while we require $E_T\ge2\gev$ for a cell to act as a jet initiator. All the partons within $\Delta R$=0.4 from the jet initiator cell are included in the formation of the jet, and we require  $E_T \ge 20 \gev $ for  a cell group to be considered a jet.
\eit

\begin{figure}[thb]
\centering
\centerline{\includegraphics[height=5.6cm]{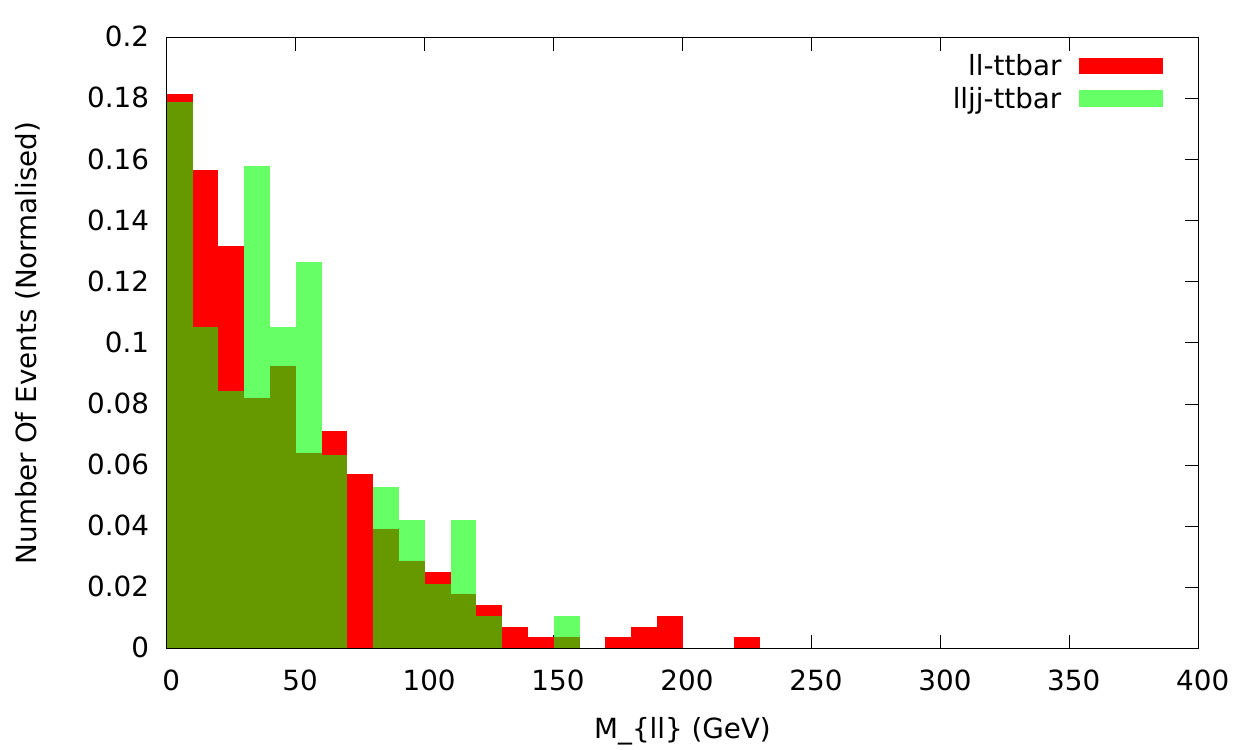}
\hskip 10pt \includegraphics[height=5.6cm]{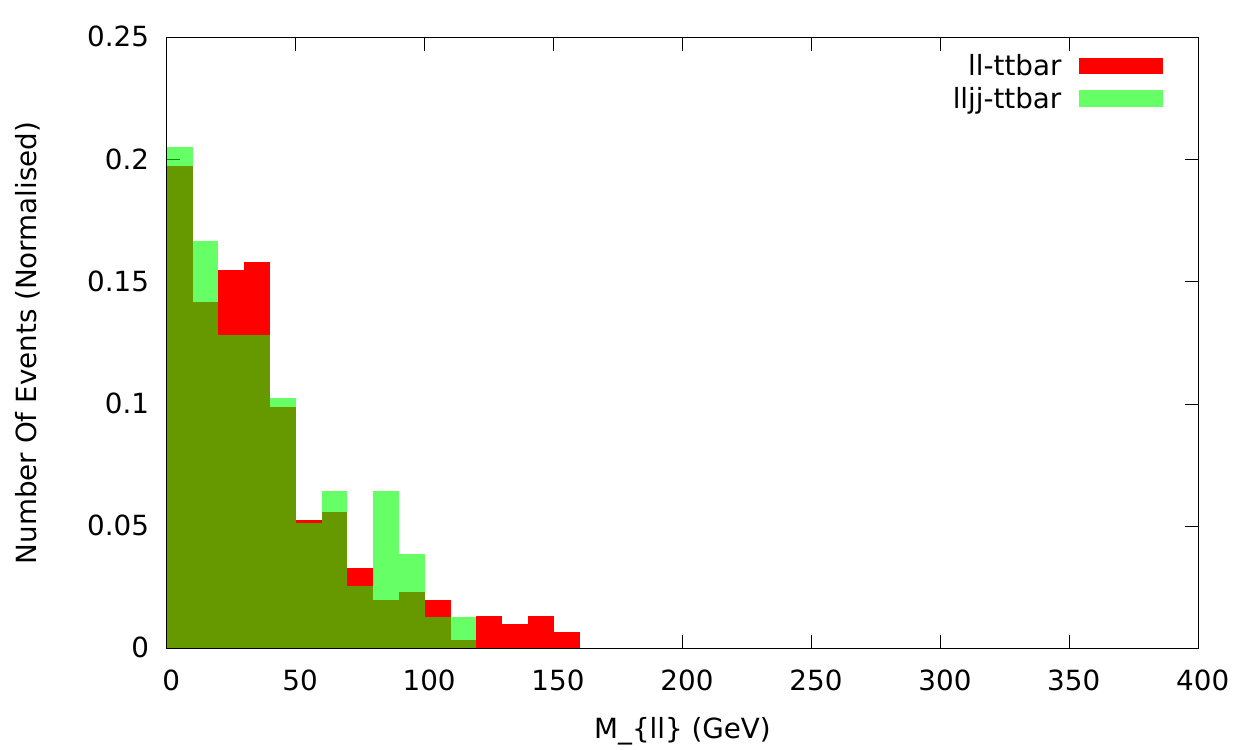}}
\caption{$M_{\ell\ell}$ distribution of the leptons for $t\bar{t}$ events at LHC with $E_{cm}= 14$ TeV in left $E_{cm}= 8$ TeV in right. The distributions for $\ell\ell$ and $\ell\ell2j$ events are shown in red and light green respectively (dark green regions correspond to the overlap of the two distributions).}
\label{fig:back-minv}
\end{figure}

\begin{figure}[thb]
\centering
\centerline{\includegraphics[height=5.6cm]{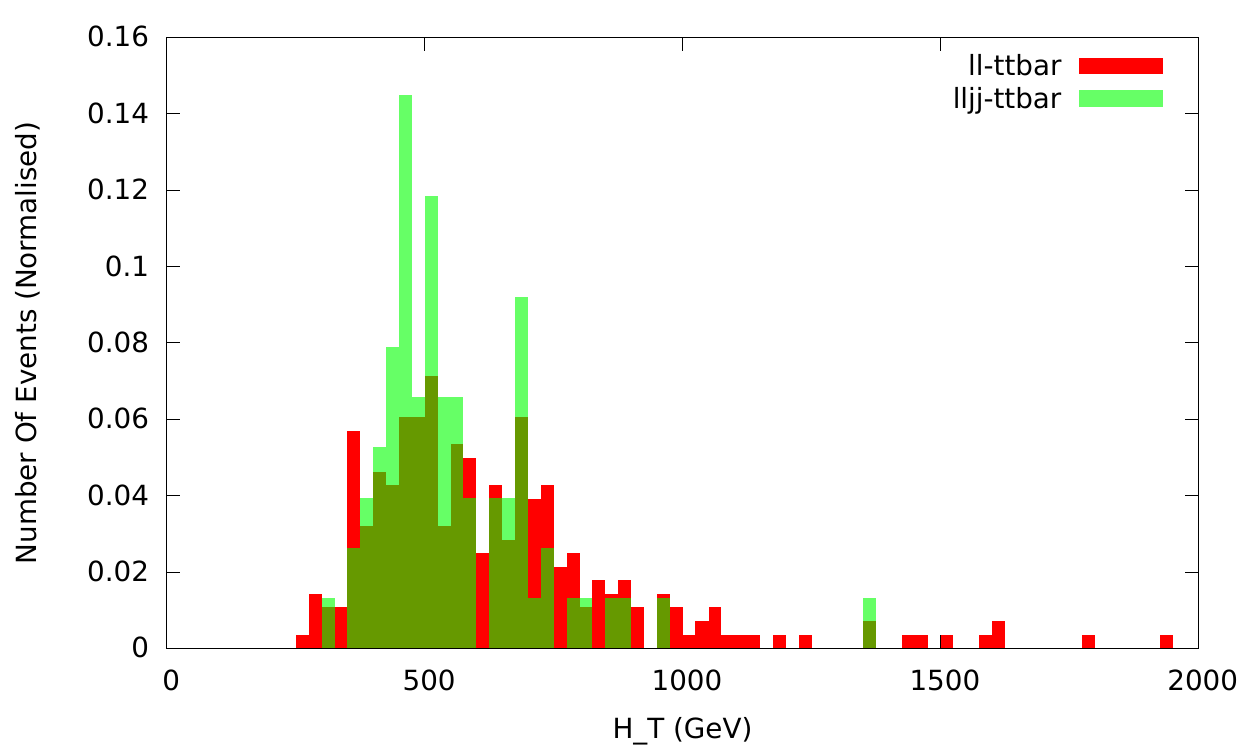}
\hskip 10pt \includegraphics[height=5.6cm]{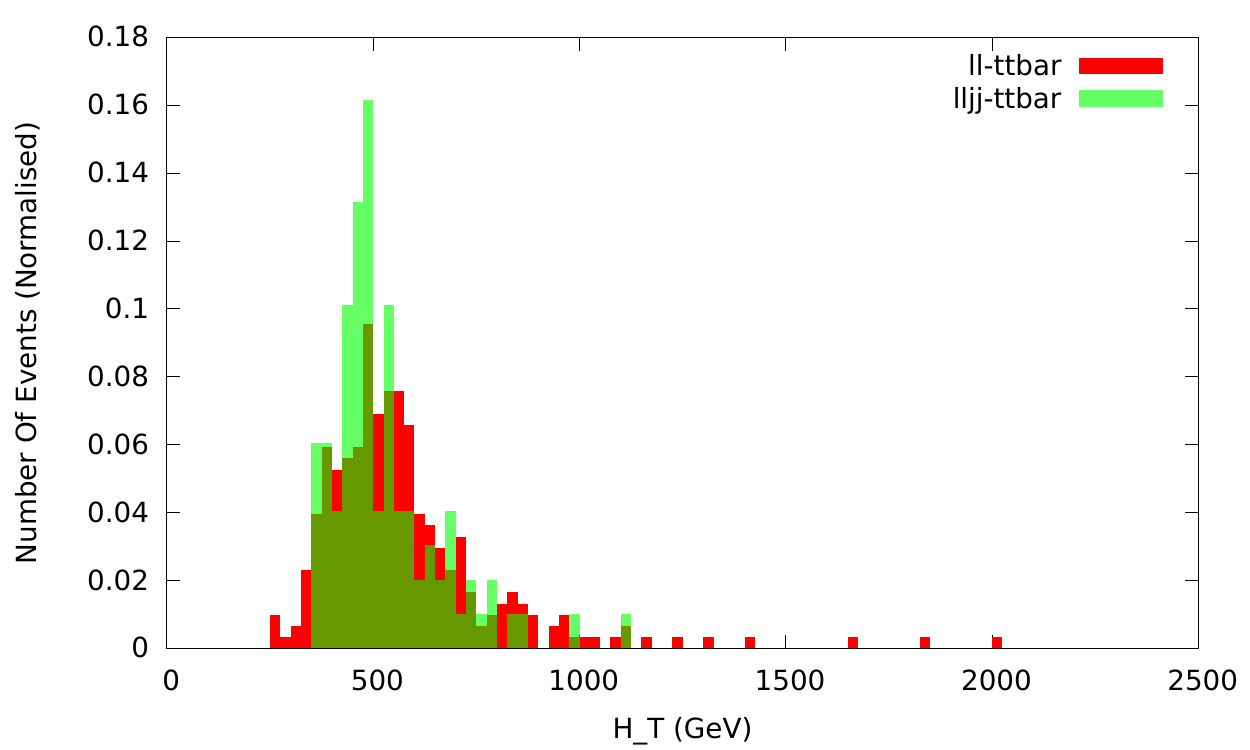}}
\caption{$H_T$ for $t\bar{t}$ events at LHC with $E_{cm}= 14$ TeV in left $E_{cm}= 8$ TeV in right. The distributions for $\ell\ell$ and $\ell\ell2j$ events are shown in red and light green respectively (dark green regions correspond to the overlap of the two distributions).}
\label{fig:back-HT}
\end{figure}

We now define the invariant lepton mass $M_{\ell\ell}$ and the transverse event mass $H_T$ by
\beq
M^2_{\ell\ell} =  2|{\bf p_1}||{\bf p_2}| \left(1 - \cos\theta \right) \,, \qquad H_T=\sum_{\ell,j} |\pp_\perp| \,,
\eeq
where $ \theta $ is the relative angle of the lepton momenta ${\bf p_1}$ and ${\bf p_2}$, and we neglected the lepton masses; $ \pp_\perp$ denotes the  corresponding momenta perpendicular to the beam. The $M_{\ell\ell}$ and $H_T$ distributions are plotted in Fig.~(\ref{fig:sig-inv-mass}) for the signal subprocesses $\bar{e} \bar{e}jj$ (lilac) and  $eejj$ (green). We see that the signal distribution from $\bar{e} \bar{e}jj$ peaks at values $ M_{\ell\ell} \sim 1 -2 \tev $ (this property is independent of $ \Lambda $), while the SM background is already much suppressed at $ M_{\ell\ell} \simeq 2 ~\tev$ (see Fig.~(\ref{fig:back-minv})). Also from Fig.~(\ref{fig:sig-inv-mass}) we find that the corresponding $H_T$ distribution for signal events peaks at $ 500-1000~\gev $, while the background is already negligible at $ 1 ~\tev $ (see Fig.~(\ref{fig:back-HT})).

The signal events consists of leptons plus jets and do not involve particles like neutrinos that are not detected at the LHC, while the background events that closely mimic the chosen signal, do produce neutrinos which passes through the detector undetected. Therefore, signal events will be characterized by having zero missing transverse to the beam, while any non-zero measurement of the following missing transverse energy
\beq
E_{\not T} = \left| \sum_{\ell,j} \pp_\perp \right|
\eeq
(where the vector sum is over all visible leptons and jets, and $ \pp_\perp$ denotes the  corresponding momenta perpendicular to the beam) will indicate the presence of particles like neutrinos in the final state, and will correspond to a background event~\footnote{Of course, real detectors miss particles, and may misidentify one type of event for the other.}. We will then require $E_{\not T}\le 15~\gev$, in addition to $C1$ in Eq.~(\ref{eq:c1}). The usefulness of this cut can be gauged by comparing the cross sections associated with $t\bar t$ production listed in table \ref{tab:ttbar}: $ \sigma_{\ell\ell} $ denotes the dilepton production cross section (two same-sign leptons plus anything); $\sigma_{\ell\ell jj}$ the dilepton plus two jet cross section, when both jets have transverse momentum $ p_T\up j \ge 25 ~\gev$; $ \sigma_{\ell\ell jj -0}$ the dilepton plus 2 jet cross section~\footnote{Our simulations generated no $t\bar t $ events with $ E_{\not T} =0 $; the corresponding limits for $\sigma_{\ell \ell j j - 0}$, are obtained as follows: if a simulation with $N$ events produces less than 1 event for a process with cross section $ \sigma $, then $ L \sigma/N < 1$, where $L$ is the luminosity, so that $ \sigma < N/L$. The numbers presented were obtained using $L=10~$ fb$^{-1}$ at $8$ TeV and $L=1~$ fb$^{-1}$ at $14$ TeV.} with zero missing energy ($E_{\not T} =0 $) and $p_T\up j\ge 25~\gev $ and for which we obtain less than one event in our simulations. Finally, $ \sigma_{\ell\ell jj -15}$ the dilepton plus 2 jet cross section with missing energy $E_{\not T} \le 15$~\gev and $p_T\up j\ge ~25~\gev $. We will use these last two quantities to derive a bound on $ \Lambda $ (for $E_{CM}=8~\tev $), and determine the expected sensitivity for $E_{CM}=14~\tev $. 

\begin{table}[ht]
\begin{center}
\begin{tabular}{|c|c|c|c|c|}
\hline
\hline
 $t\bar{t}$ Production at LHC & $\sigma_{\ell\ell}\up{tt}$ & $\sigma_{\ell\ell jj}\up{tt}$ & $\sigma_{\ell\ell jj-0}\up{tt}$ & $\sigma_{\ell\ell jj-15}\up{tt}$\\
\hline
$E_{CM}$=14 TeV & 0.47 & 0.145 & $\le 0.85 \times 10^{-3}$ & 0.0077\\
\hline
$E_{CM}$=8 TeV & 0.089 & 0.0282 & $\le 0.162 \times 10^{-3}$ & 0.001\\
\hline
\hline
\end {tabular}
\end{center}
\caption {Dominant SM background cross-section from $t\bar{t}$ production (in pb) for same-sign dilepton events at $E_{cm}=14,\, 8$ TeV including muons, antimuons, electrons or positron. The fourth (last) column cross sections are obtained requiring $ E_{\not T} =0 $ ($E_{\not T}\le15~\gev$); see text. The cross-sections are obtained after multiplying by the appropriate $K$-factor: 2.20 for 14 TeV and 1.7 for 8 TeV.}
\label{tab:ttbar}
\end{table}

\begin{figure}[thb]
\centering
\centerline{\includegraphics[height=5.6cm]{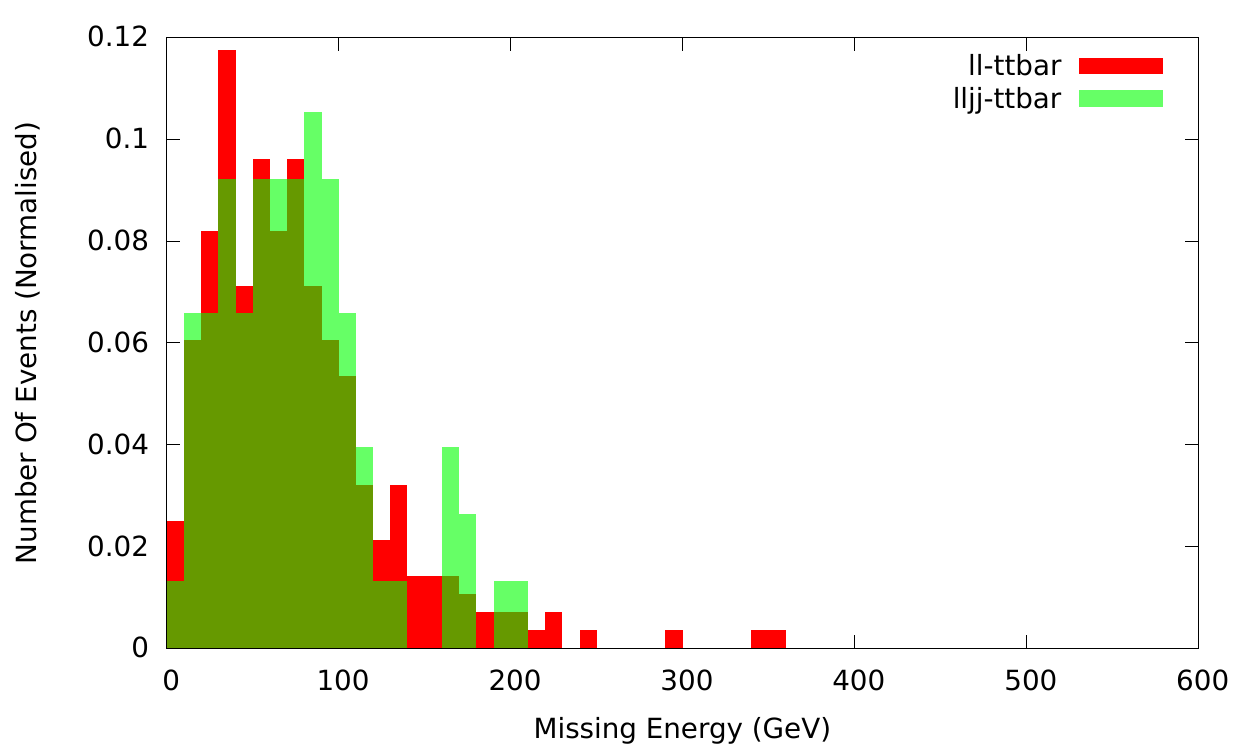}
\hskip 10pt \includegraphics[height=5.6cm]{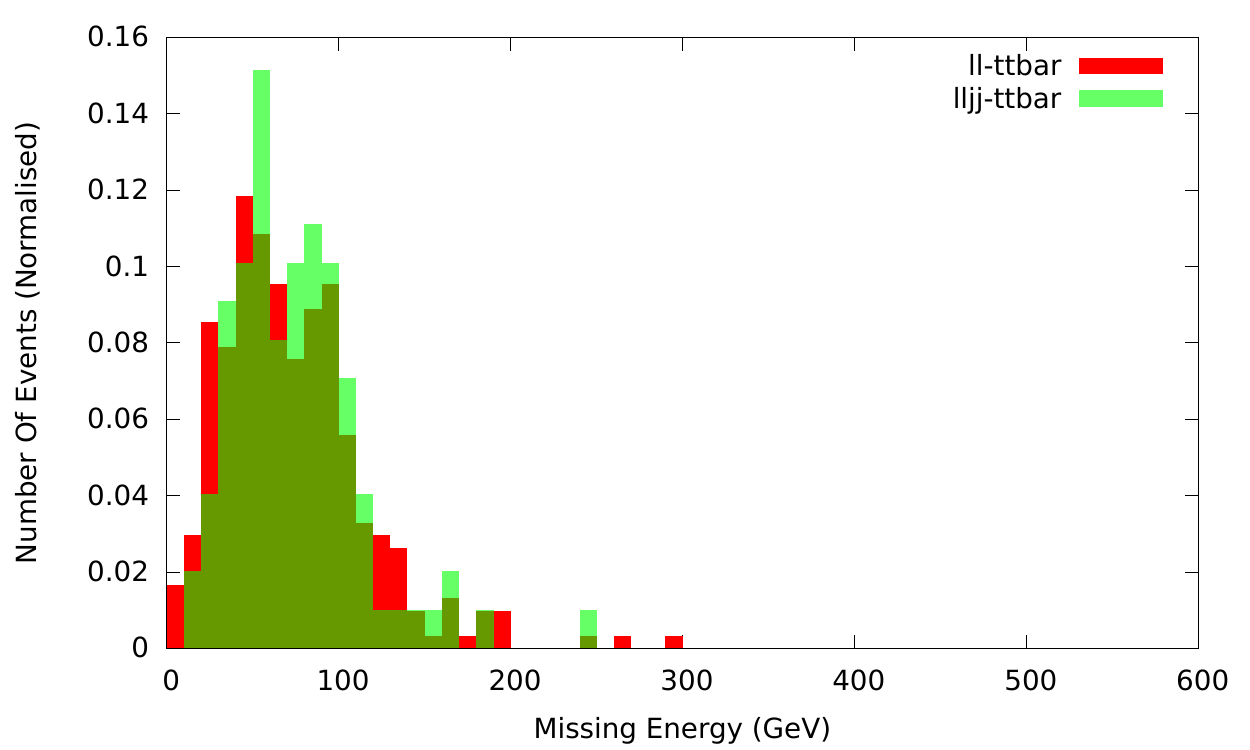}}
\caption{$E_{\not T} $ distribution for $t\bar{t}$ events at the LHC with two leptons (light green) and two leptons and two jets (red); dark green regions correspond to the overlap of the two distributions. Left: $E_{cm}= 14$ TeV; right: $E_{cm}= 8$ TeV (right). }
\label{fig:back-MET}
\end{figure}

 The missing energy $E_{\not T}$ distribution for the SM background (Fig.~(\ref{fig:back-MET})) shows that there is a very large number of $\ell\ell j j$ events with $E_{\not T}<100$ GeV, and that this is drastically reduced to a vanishingly small number when the cut $E_{\not T}<15$ GeV is imposed, as also shown in table~\ref{tab:ttbar}. This clearly indicates that with such a selection criteria, the signal events are retained (as ideally they are characterized by having zero missing energy), while the background can be reduced significantly allowing for a much improved discovery limit. In addition, the invariant lepton mass $M_{\ell\ell}$ (Fig.~(\ref{fig:back-minv})) and transverse event mass $H_T$ (Fig.~(\ref{fig:back-HT})) distributions show a characteristic difference between signal and background. For example, $M_{\ell\ell}$ distribution for signal (upper panel of Fig.~(\ref{fig:sig-inv-mass})) peaks $M_{\ell\ell} \sim 800$ GeV while the one for background (Fig.~(\ref{fig:back-minv})) peaks between $M_{\ell\ell} \sim 0-10$ GeV which can be also used for distinguishing between the contributions from new physics and the SM background.

\subsection{Bound on $\Lambda$  from LHC data} 
The LHC has not seen even a 1$\sigma$ excess from the background events so that, assuming Gaussian statistics
\begin{equation}
\nonumber
S/\sqrt{B} \le 1\,,
\end{equation}  
where $S,\,B$ denote the number of  signal and background events respectively. In terms of the luminosity $L$ and the corresponding cross sections $S=\sigma_{\rm signal}  L,\,  B=\sigma_{\rm back.}  L$ where $L$ is integrated luminosity. To date CMS has analyzed LHC data for same sign dileptons for $L=$ 19.5 fb$^{-1}$ at $8~\tev $ \cite{CMS-8TeV}; the background cross-section corresponds to $ \sigma_{\rm back.} = \sigma_{\ell\ell j j -15}\up{tt} = 1$ fb in table \ref{tab:ttbar}, we then find
\begin{equation}
\nonumber
\sigma_{\rm signal} \le 0.224 ~{\rm fb}\,.
\end{equation}
  
This in turn puts the bound on the new physics scale: we know $\sigma_{\rm signal}=273$ fb (after cuts) when  $\Lambda=100$ GeV (see table \ref{table:eeqq1}), and that the signal cross section scales as $ 1/\Lambda^6 $ (see Eq. (\ref{eq:scale})). It follows that 
 
\beq
\Lambda \ge \left( \frac{273\, {\rm fb}}{0.224\, {\rm fb}} \right)^{1/6} \times 100~ \gev = 327~ \gev\,.
\label{eq:clim}
\eeq

\subsection{Discovery limit for $\Lambda $ for $\mathcal{O}_{\ell\ell}$ at the LHC} 

We will follow the same approach to evaluate the discovery limit at the  $14~\tev$ LHC, requiring now a 3$\sigma$ excess over the background 
\begin{equation}
\nonumber
\sigma_{\rm signal} \ge 3 \sqrt{\frac{\sigma_{\rm back.}}{L}}\,.
\end{equation}
From table~\ref{tab:ttbar} we find $\sigma_B= \sigma\up{tt}_{\ell\ell j j -15} = 7.7 {\rm fb} $ while from table~\ref{table:eeqq1} we find $\sigma_{\rm signal}=2590$ fb (after cuts) when the new physics scale $\Lambda=100$ GeV. Using again the simple scaling of the signal cross section with $ \Lambda $ (see Eq.~(\ref{eq:scale})) we find 
\beq
\Lambda \le 382~ \gev \quad (L=100\,{\rm fb}^{-1},~ E_{\not T} \le 15~ \gev)\,,
\eeq
for an integrated luminosity $L=100$~fb$^{-1}$.

If we  use a stronger missing energy cut the limit improves. If we require no missing transverse energy then the background cross section drops to $ \sigma\up{tt}_{\ell \ell j j - 0} = \sigma_{\rm back.} =0.85 $ fb (see table~\ref{tab:ttbar}) so that, for the same luminosity,
\beq
\Lambda \le 459 ~\gev \quad (L=100\,{\rm fb}^{-1},~ E_{\not T} =0~ \gev)\,.
\label{eq:flim}
\eeq

\subsection{Hadronically-quiet trilepton at the LHC}

\begin{figure}[thb]
\centering
\centerline{\includegraphics[height=5.5cm]{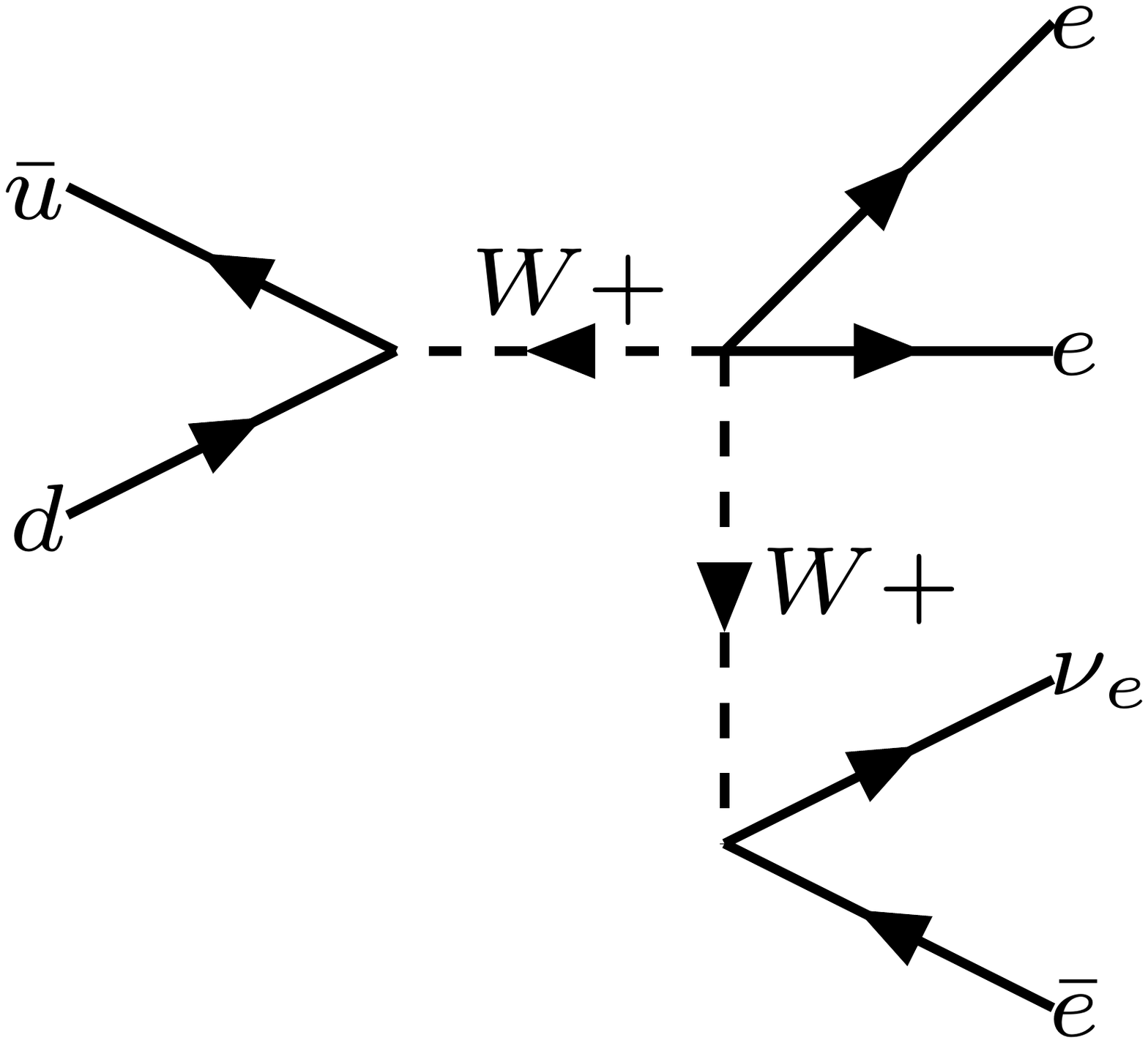}
\hskip 10pt \includegraphics[height=5.5cm]{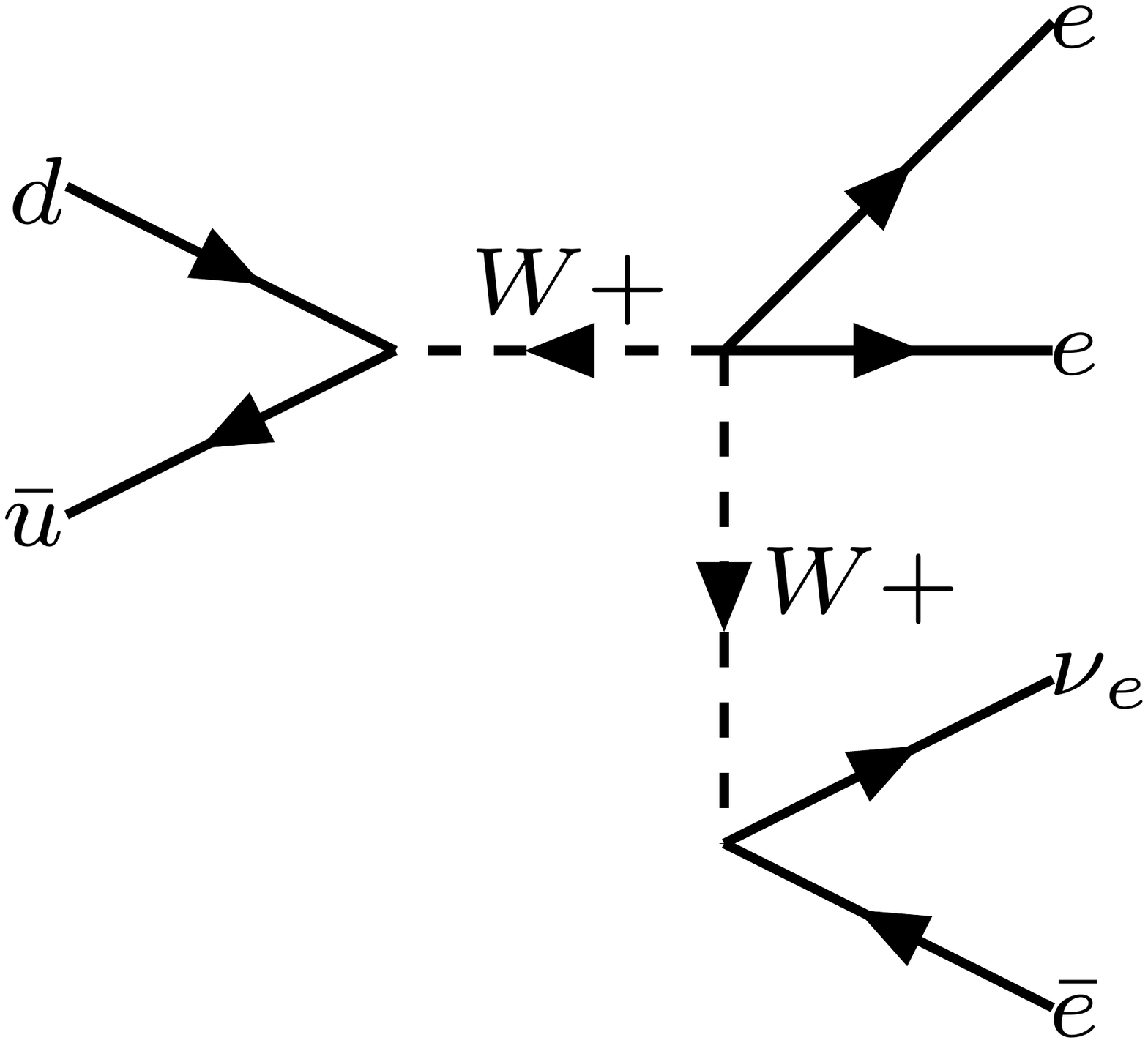}}
\caption{Feynman diagrams generated by the operator $\ocal_{\ell\ell}$ and which contribute to the hadronically-quiet trilepton channel $pp \to ee\bar{e}\nu_e$.}
\label{fig:back-SPT}
\end{figure}

The  operator $\mathcal{O}_{\ell\ell}$ can also produce hadronically-quiet trilepton events at the LHC, which is sometimes favoured as a signal for new physics because of its clean signature and small SM backgrounds. We find however, that for $\mathcal{O}_{\ell\ell} $ the total signal cross-section is only $3.28$ fb for $\Lambda=$ 100 GeV (see table~\ref{tab:3l-sig}). This small value  results from the relatively small branching ratio of the $W$ boson into leptons and the absence of a $t$-channel diagram that generated an important contribution to the $ \ell\ell j j $ final state (compare Fig.~(\ref{fig:back-SPT}) and Fig.~(\ref{fig:fd12})).

\begin{table}[ht]
\begin{center}
\begin{tabular}{|c|c|c|c|}
\hline
\hline
 $p,p\to \ell,\ell,\ell,\nu_{\ell}$ & $\sigma$ (fb) & $\sigma_1$ (fb) & $\sigma_2$ (fb) \\
\hline
$p,p\to \bar{e},\bar{e},e,\bar{\nu_e}$ & 1.2 & 0.70 & 0.695\\
\hline
$p,p\to e,e,\bar{e},\nu_e$ & 0.44 & 0.39 & 0.385\\
\hline
Total (including $\mu$) & 3.28 & 2.18 & 2.16\\
\hline
\hline
\end {tabular}
\end{center}
\caption {Cross-sections for the hadronically-quiet trilepton signal at the LHC with $E_{cm}=$14 TeV. $\sigma_1$ is obtained by imposing the cut $C1$ (cf. Eq.~(\ref{eq:c1})), and $\sigma_2$ by imposing both  $C1$ and demanding $ |M_{e\bar{e}} - M_z |>15~ \gev$.}
\label{tab:3l-sig}
\end{table}

\begin{table}[ht]
\begin{center}
\begin{tabular}{|c|c|c|c|c|}
\hline
\hline
 SM Production at LHC & $\sigma_{\ell\ell\ell}$ (pb) & $\sigma_{\ell\ell \ell-M_{\ell\ell}}$ (pb) & $\sigma_{\ell\ell \ell-M_{\ell\ell}-0}$ (fb) & $\sigma_{\ell\ell \ell-M_{\ell\ell}-15}$ (fb)\\
\hline
 $t\bar t$& 0.99 & 0.031 & $ \le$ 0.85& 1.69 \\
\hline
$WZ$ & 0.73 & 0.171 & $\le$ 0.058   & 0.46\\
\hline
\hline
\end {tabular}
\end{center}
\caption {Dominant SM background cross-sections $\sigma_{\ell\ell\ell\ell} $ (in pb)  generated by $t\bar{t}$ and $WZ$ production for hadronically-quiet trilepton events at $E_{cm}=14$ TeV including $\mu^\pm$ and $ e^\pm$; the third column provides the numbers when the cut $ | M_{\ell\ell} - M_z| > 15 \gev $ is imposed on the events. The cross-sections in the fourth and fifth columns are obtained requiring $ E_{\not T} =0 $ and $E_{\not T}\le15\gev$ respectively (see text for details). The cross-sections are obtained after multiplying by the appropriate $K$-factor (=2.20 for $t\bar{t}$).}
\label{tab:3l-back}
\end{table}

$WZ$ production generates a significant background to the process being considered; in order to suppress it we require that the invariant mass $ M_{\ell\ell} $ between any opposite-sign leptons of the same flavor satisfies $ | M_{\ell\ell} - M_z| > 15 ~\gev $. Comparing the results in table~\ref{tab:3l-sig} and table~\ref{tab:3l-back} we see that using the cuts we are able to significantly reduce the background compared to the signal cross section of $2.16 $ fb for $ \Lambda = 100~ \gev $. Following a procedure similar to the one we used in analyzing the same-sign dilepton events and using Eq.~(\ref{eq:scale}) we determined the sensitivity to $ \Lambda $ using this hadronically-quiet trilepton channel; using a $3\sigma$ limit we obtain, 
\beq
\Lambda \le 130 ~\gev \quad ( E_{\not T} \le 15 ~\gev)\,, \qquad \Lambda \le 140 ~\gev \quad (E_{\not T} = 0 ~\gev)\,,
\eeq
for a luminosity of $ 100 ~{\rm fb}^{-1} $. As expected from the smaller cross section these limits are weaker than the ones previously obtained.

\subsection{Applicability of the effective field theory}
\label{sec:appl}

\begin{figure}[thb]
\vspace{1cm}
$$
\includegraphics[height=3cm]{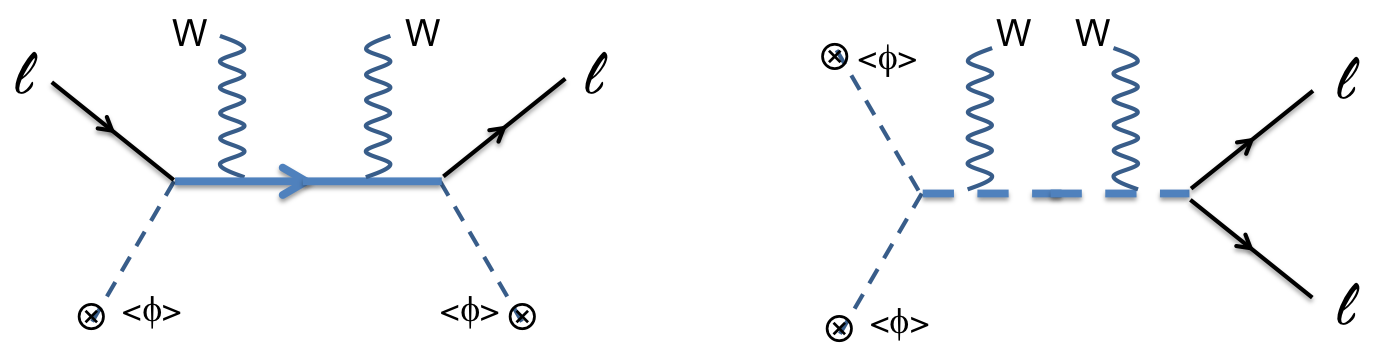}
%\vspace{-1cm}
$$
\caption{Tree-level graphs that can generate the operators $ \ocal_1$ and $ \ocal_3 $ by the exchange of a heavy fermion (left) or scalar (right), both isotriplets with unit hypercharge (denoted by the thick internal lines).}
\label{fig:PTG_graphs}
\end{figure}

The results above indicate that the reach on new physics scale $\Lambda$ lies in the $\simeq 500 ~ \gev$ for the same sign dilepton signature generated by the operator(s) $\ocal_{\ell\ell}$ at the LHC with a $ 14$ TeV CM energy. On the other hand, the dilepton invariant mass distribution (Fig.~(\ref{fig:sig-inv-mass})) peaks at $\sim 1 $ TeV, a significantly higher value. We must then investigate whether the effective theory approach is valid in the above processes. Specifically, within the paradigm we have adopted (that of a weakly coupled, renormalization and decoupling heavy physics), we must determine whether this implies that in the reactions under consideration one or more heavy particles carries momentum $ \ge \Lambda$ (in which case they could be directly produced and the effective approach would not be applicable). In order to investigate this we display in Fig.~(\ref{fig:PTG_graphs}) the types of heavy physics that could generate at tree level the vertex of interest (containing two charged leptons of the same sign, two $W$ bosons of the same sign and two scalar isodoublets).
 
\begin{figure}[thb]
\centering
\centerline{\includegraphics[height=7cm]{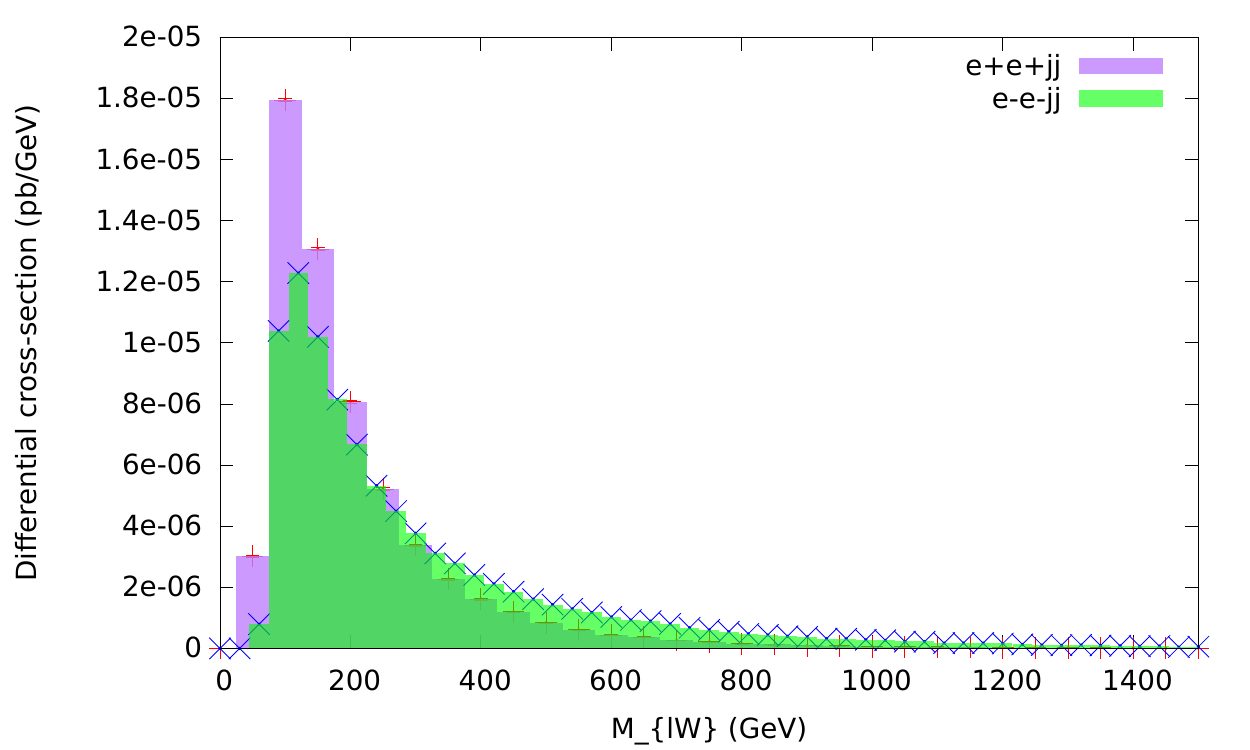}}
\caption{Invariant mass distribution of the lepton with jets at LHC with $E_{cm}= 14$ TeV. The red dots (on the lilac histogram) correspond to $\bar{e}\bar{e}jj$ final state; while the blue crosses (on the green histogram) correspond to $eejj$ final state (the green area corresponds to regions where these distributions overlap).}
\label{fig:minv-EW1}
\end{figure}

We see from Fig.~(\ref{fig:PTG_graphs}) that there are two possible cases
\ben
\item When the heavy particles carry a momentum equal to the invariant dilepton mass, in which case the heavy particle is a boson isotriplet of unit hypercharge; or,
\item When the heavy particles carry a momentum equal to the lepton-$W$ invariant mass, in which case the heavy particle $\Sigma$ will be a heavy fermion isotriplet or isosinglet of zero hypercharge. 
\een
The comments above indicate that the effective theory would not be applicable in case 1. The situation for the fermions $ \Sigma $ (case 2), however, is different: we see from Fig. \ref{fig:minv-EW1} that the lepton-$W$ invariant mass peaks at $ \lesssim 200~\gev $, significantly below the limit on $ \Lambda $. It follows that the limits we derive are applicable for the case where the new heavy physics corresponds to the same fermions $\Sigma$ associated with Type III seesaw mechanisms \cite{type3} for neutrino mass generation see the appendix \ref{sec:lg.ptg})~\footnote{Type I involves fermion isosinglets and the corresponding effective operators would not have tree-level couplings to the $W$.}.

\bigskip

Direct searches of such heavy fermion isotriplets $\Sigma $ (using the full theory) at the $7$ TeV LHC yield a bound $ m_\Sigma \gesim  180 ~\gev $ \cite{type3iv}, with which our result $ \Lambda \gesim 300 ~\gev $ compares favourably.  In addition, reference \cite{CMSdimuon} presents an analysis where dimuon + jets data from the 8 TeV LHC are used to derive constraints on heavy Majorana neutrinos that can mix with the muon-neutrino (these correspond to the case where the $\Sigma $ are isosinglets). The results exclude heavy fermion masses in the range $ 90-500~\gev$ provided the mixing is  $\ge 0.005$ (for the low mass limit) and $ \ge 0.6$ for the higher excluded mass; these constraints are then more stringent than the ones derived above for the case of large mixing. There are also several studies of the discovery potential of Type III seesaw fermions at  the $14$ TeV LHC through direct production \cite{type3i,type3ii,type3iii} in lepton-rich final states, and though a comparative signature space analysis of the reach of direct production versus effective theory is beyond the scope of this paper, a benchmark point analysis in \cite{type3i} suggests that the sensitivity to $ \Lambda $ obtained using effective field theory will be competitive to the one derived form direct searches.

\section{Constraints on PTG operators with right-handed neutrinos}
\label{sec:rhn}

In section \ref{sec:decays} we listed the operators that do not contain \rh\ neutrinos and obtained the most stringent bounds on the scale of new physics. In this section we do a similar study for operators that do contain \rh\ neutrinos. The list of such operators and their expressions in unitary gauge can be found in Appendix \ref{sec:ptg.rh}. Most of these operators contain vertices with 3 fields (see table~\ref{tab:rhn}), one of which is a $W,\, Z$ or $H$ boson; the strictest limits on $ \Lambda $ are then derived from $Z, W, H$ decays and neutrino magnetic moment, whenever kinematically allowed. When the \rh\ neutrinos are too heavy for these decays to occur the limits are weaker, as is the case for the operator $(\overline{\nu^c} \nu)|D\phi|^2 $ that does not contain a three-legged vertex; we comment on this situation at the end of this section. 

In the discussion below we will assume that the \rh\ neutrinos have a Majorana mass term of the form $ \nu_R^T C M_\nu \nu_R $ that, for simplicity, we assume to be fully degenerate: $M_\nu = m_\nu \mati$. In cases where the $W,\, Z$ or $H$ decays are allowed, the effects of mixing (generated by Dirac mass terms) will be small~\cite{delAguila:2008ir} and we will ignore them. As in section \ref{sec:decays} the limits on $ \Lambda $ obtained below are derived assuming that the operator coefficient is $O(1)$, if this is not the case such limits apply to the scale $ \tilde\Lambda$ defined in that section; we will also ignore the possibility of cancellations among various effective operator contributions.

\subsection{Operators contributing to $Z$-invisible decay}

There are  two such operators (see table~\ref{tab:rhn}):
\beq
\half (\ncb \gamma^\mu N )(i \phi^\dagger \stackrel \leftrightarrow D_\mu \phi) \supset  \frac{v^3}{2\sqrt{2}}  \overline{\nu^c} \slashed Z P_L \nu \,, \qquad
   |\phi|^2 (\ncb \sigma^\mn \nu) B_\mn  \supset -\frac{v^2 \sw}2\left( \overline{\nu^c} \sigma^\mn P_R \nu \right) Z_\mn  \,, 
\eeq
(where $ Z_\mn = \partial_\mu Z_\nu - \partial_\nu Z_\mu$) that generate the following contributions to the invisible $Z$ decay width:
\bea
\frac{v^2 \mz}{\Lambda^3}  \ncb \slashed Z P_R \nu : &&
\Gamma_{Z1}( Z \to \nu_R \nu_R )=\frac{1}{96\pi} \left(\frac{v^2}{\Lambda^3} \right)^2(\mz^2-m_\nu^2)\sqrt{\mz^2-4m_\nu^2}  \,,\cr
\frac{v \mz}{\Lambda^3}(\ncb\sigma^{\mu\nu}P_R \nu)Z_{\mu\nu}: && \Gamma_{Z2}( Z \to \nu_R \nu_R)=  \frac{1}{12\pi} \left( \frac{\mz v}{\Lambda^3} \right)^2  \left( \mz^2 + 2  m_\nu^2  \right) \sqrt{\mz^2 - 4 m_\nu^2 }\,;
\label{eq:Z-decay1}
\eea
(we made an $O(1)$ change in the operator coefficients in order to have a uniform  normalization of the three-legged vertices). 

\begin{figure}[thb]
$$
\includegraphics[height=7.0cm]{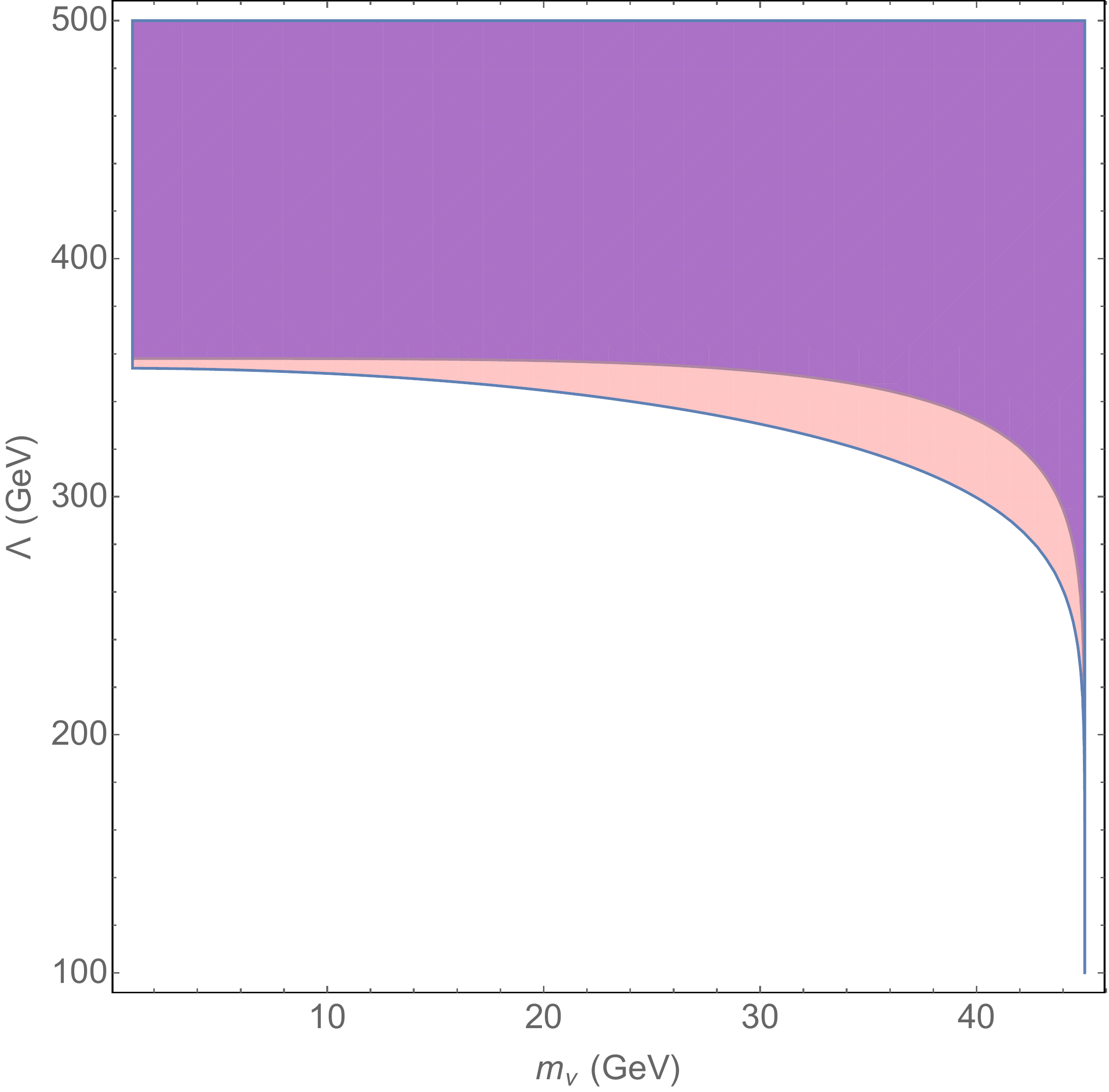}
$$
\caption{$Z$ invisible decay limit on neutrino mass $m_\nu$ versus new physics scale $\Lambda$ plane derived from Eq.~(\ref{eq:Z-decay1}). The  light red region and above is allowed by operator $ \ncb\slashed{Z}P_L \nu$, and the lilac region is allowed by $(\ncb \sigma^{\mu\nu}P_{L,R} \nu)Z_{\mu\nu}$.}
\label{fig:Z1}
\end{figure}

The invisible $Z$-decay width in the standard model is $\Gamma(Z \to{\rm inv.})= 499 \pm 1.5 $MeV~\cite{PDG}. If the $Z$ decay to right handed neutrinos is kinematically allowed, then, at $3\sigma$, $ \Gamma_{Z1,Z2} < 4.5$ MeV that gives the contours plotted in Fig.~(\ref{fig:Z1}). We see from this that the strictest limits on $ \Lambda $ are obtained when the \rh\ neutrino mass is small:
\beq
Z1( m_\nu \ll \mz): ~\Lambda > 354 \gev\,, \qquad
Z2( m_\nu \ll \mz): ~\Lambda > 358 \gev\,.
\eeq

\subsection{Operators  with \rh\ neutrinos contributing to $H$-invisible decay}

\begin{figure}[thb]
$$
\includegraphics[height=7.0cm]{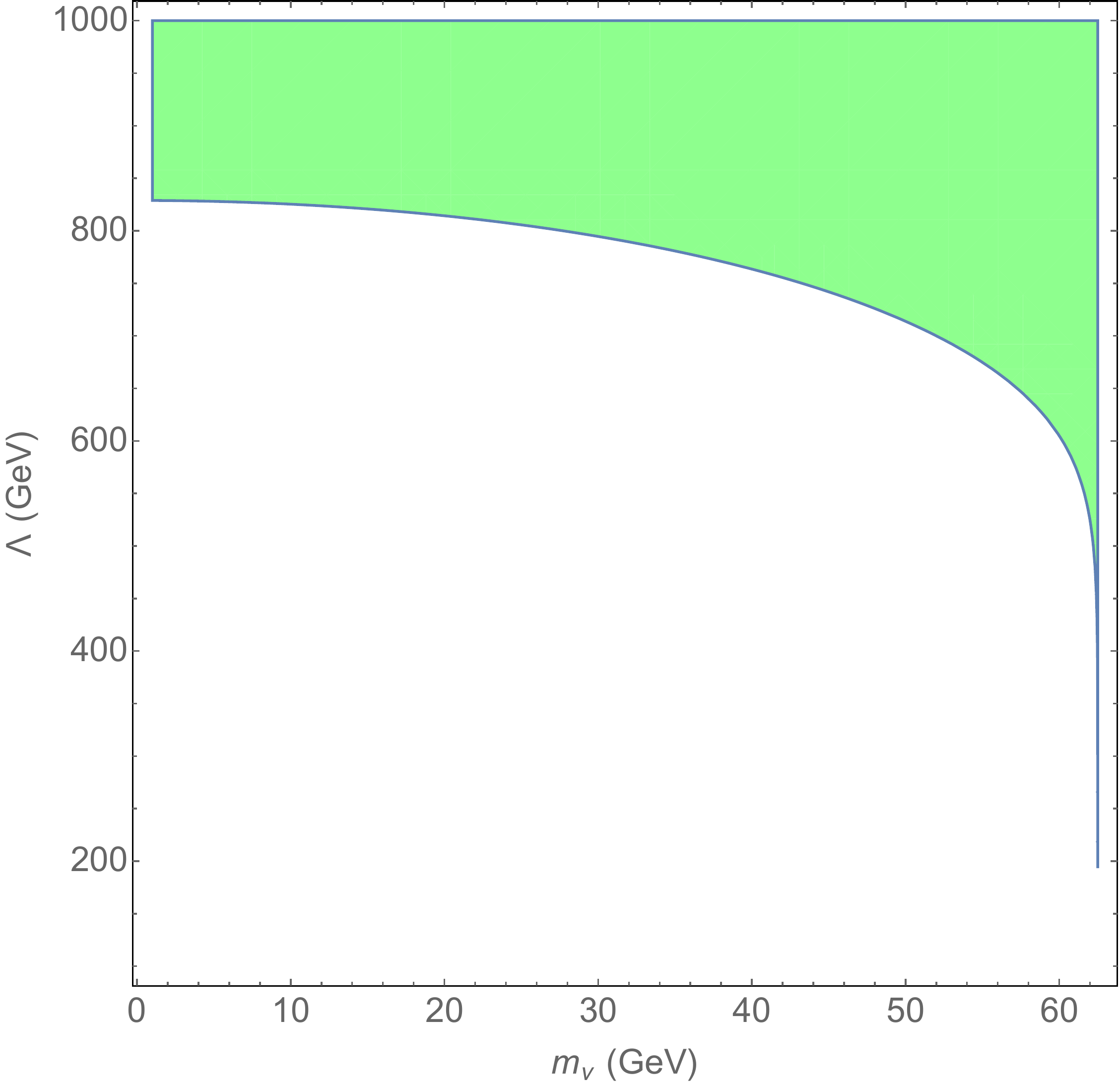}
$$
\caption{H invisible decay limit on neutrino mass $m_\nu$ versus new physics scale $\Lambda$ plane derived from Eq.~(\ref{eq:Hbound}). The shaded region is allowed by operator $\ncb\nu |\phi|^4$.}
\label{fig:H1}
\end{figure}

Again referring to table~\ref{tab:rhn} we see that there  a single operator of this type:
\beq
\ncb\nu |\phi|^4 \supset v^3 H\left( \overline{\nu^c} P_R \nu \right)\,,
\eeq
which yields the following invisible decay width
\beq
\Gamma(H \to \nu_R \nu_R)=\frac{1}{16\pi}{\mh (v/\Lambda)^6 }\left(1 - \frac{2 m_\nu^2}{\mh^2} \right) \sqrt{1 - \frac{4 m_\nu^2}{\mh^2}} \,.
\label{eq:Hbound}
\eeq
Using the limit  ${\rm Br}( H \to {\rm inv}) < 0.3$~\cite{invisible_higgs2}, and $\Gamma_H^{\rm SM}=4 $MeV for the total SM contribution to the $H$ width, we find $ \Gamma(H \to {\rm inv}) < 1.7 \mev$, which we plot in  Fig.~(\ref{fig:H1}). For light neutrinos this implies, 
\beq
\Lambda> ~828 \gev\,, \quad (m_\nu \ll \mh )\,.
\eeq

\subsection{Operators with \rh\ neutrinos contributing to $W$-leptonic decay}

There are 3 such operators (see table~\ref{tab:rhn}):
\bea
\half (\ncb \gamma^\mu E ) \left(\phi \epsilon \stackrel \leftrightarrow D_\mu \phi \right) &\supset& - i \mw v^2 \ncb \slashed W^+ P_L e \,,\cr
  (\ncb \sigma^\mn e)(\phi\eps\tau^I \phi) W^I_\mn  &\supset& \sqrt{2} v^2 \left( \partial_\mu W_\nu^+ \right) \left( \ncb  \sigma^\mn P_R e \right) \,,\cr 
  (\ncb D_\mu e)(\phi\eps D^\mu \phi)  &\supset&  -i \sqrt{2}\, \mw v  W^+_\mu \left( \ncb \partial^\mu P_R e \right)\,,
  \eea
that leads to three contributions to the leptionic $W$ decay width:
\bea
\frac{\mw v^2}{\Lambda^3}\ncb \slashed W^+ P_L e: && \Gamma_{W1}= \frac{1}{192\pi} \left(\frac{v^4}{\Lambda^6\mw^3} \right) (\mw^2-m_\nu^2)^2(2\mw^2+m_\nu^2)\,,\cr
\frac{\mw v}{\Lambda^3}\left( \partial_\mu W_\nu^+ \right) \left( \ncb  \sigma^\mn P_R e \right):&& \Gamma_{W2}=\frac{1}{12\pi} \left(\frac{v^2}{\Lambda^6 \mw} \right) (\mw^2-m_\nu^2)^2 (\mw^2+2 m_\nu^2)\,,\cr
\frac{\mw v}{\Lambda^3} {W^+_\mu \left( \ncb \partial^\mu P_R e \right)}:&& \Gamma_{W3}=\frac{1}{192\pi} \left(\frac{v^2}{\Lambda^6\mw^3} \right) (\mw^2-m_\nu^2)^4\,.
\label{eq:Wbound}
\eea

\begin{figure}[thb]
$$
\includegraphics[height=7.0cm]{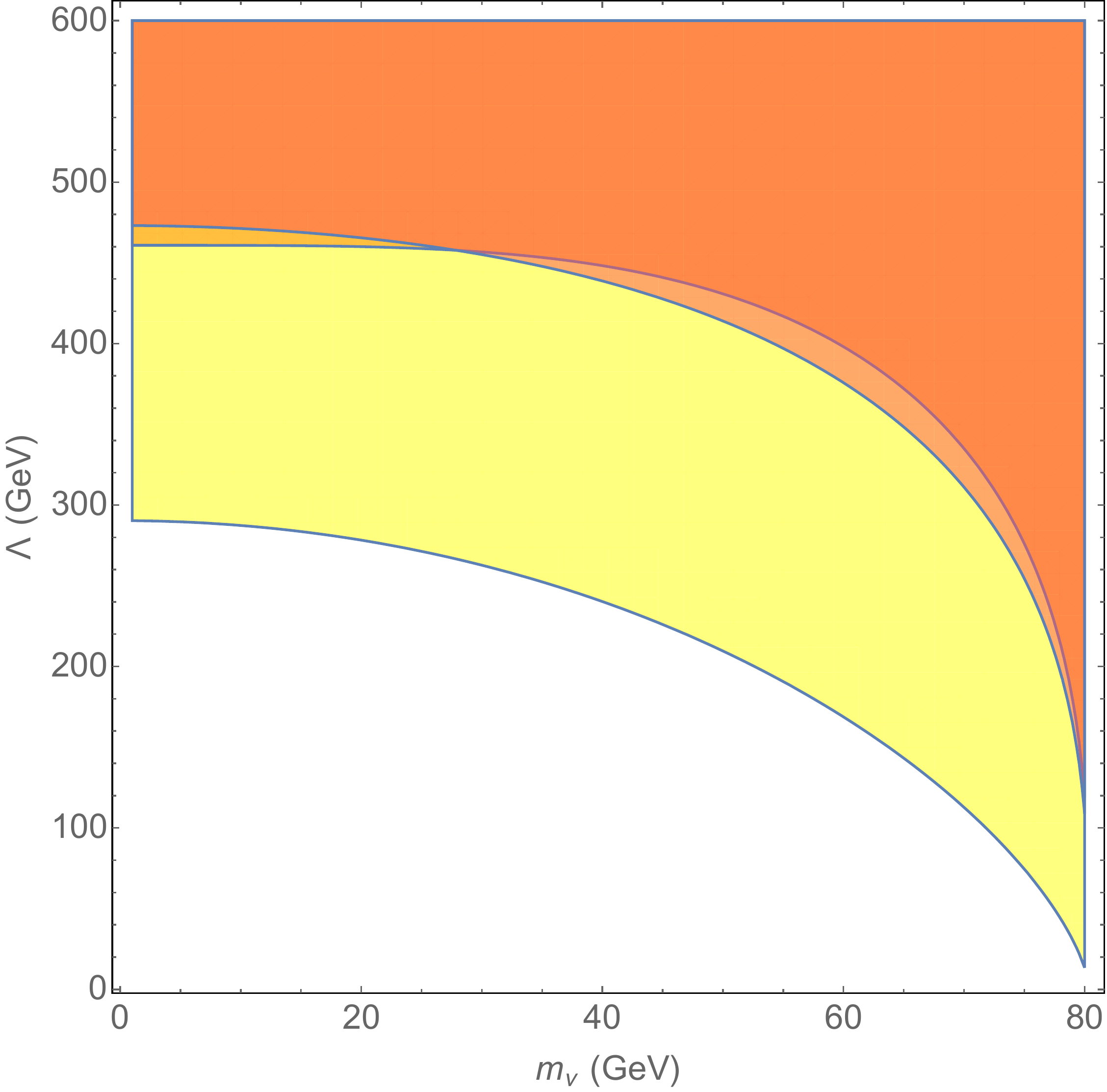}
$$
\caption{$W$- decay limit on neutrino mass $m_\nu$ versus new physics scale $\Lambda$ plane derived form Eq.~(\ref{eq:Wbound}). The yellow region and above is allowed by $W^+_\mu \left( \ncb \partial^\mu P_R e \right)$, the darker orange region and above is allowed by $(\ncb\sigma^{\mu\nu}P_R e)\partial_\mu W^{+}_\nu$, and the orange region and above is allowed by $  \ncb \slashed W^+ P_L e$.}
\label{fig:W1}
\end{figure}

 Now, the branching fraction of $W$ to $\ell+\nu_\ell$ (combined leptonic final states) is $ \Gamma( W \to \ell \nu_\ell)/\Gamma_W = (10.86\pm0.09)\%$ with $\Gamma_W= 2.085\, \gev $ \cite{PDG}. This produces the contours of of Fig.~(\ref{fig:W1}) in 3$\sigma$ limit; for light neutrinos
\bea
W1 ~(m_\nu \ll \mw):&& \Lambda > 473 \gev\,, \cr
W2~(m_\nu \ll \mw):&& \Lambda> 460 \gev\,, \cr
W3~(m_\nu \ll \mw):&& \Lambda>  290\gev  \,.
\eea

\subsection{Operators contributing to the neutrino magnetic moment}
The operators relevant for this constraint are listed in Eq.~(\ref{eq:3lv}), for which the most significant constraint is derived from the cooling rate of red-giants and other astrophysical objects~\cite{plasmon}. Using the results in \cite{nuSM-5} we find that the strongest restriction is obtained by requiring that the process $ \gamma + \nu_{\rm light} \to \nu_{\rm heavy}$ is not a very efficient cooling mechanism in supernovae \cite{plasmon}:
\beq
\Lambda >  \left( 1 - \frac{4 m_\nu^2}{\omega_P^2} \right)^{1/4}\!\!\times 47 \,\tev \,; \quad ~2 m_\nu \le  \omega_P\simeq 30\, \mev\,,
\label{eq:mnubound}
\eeq
where $ \omega_P$  the typical supernova plasma frequency; this same limit is often expressed by the constraint that the $ \nu$ magnetic moment is smaller than $ 3 \times 10^{-12} \mu_B$.

\begin{figure}[thb]
$$
\includegraphics[height=7.0cm]{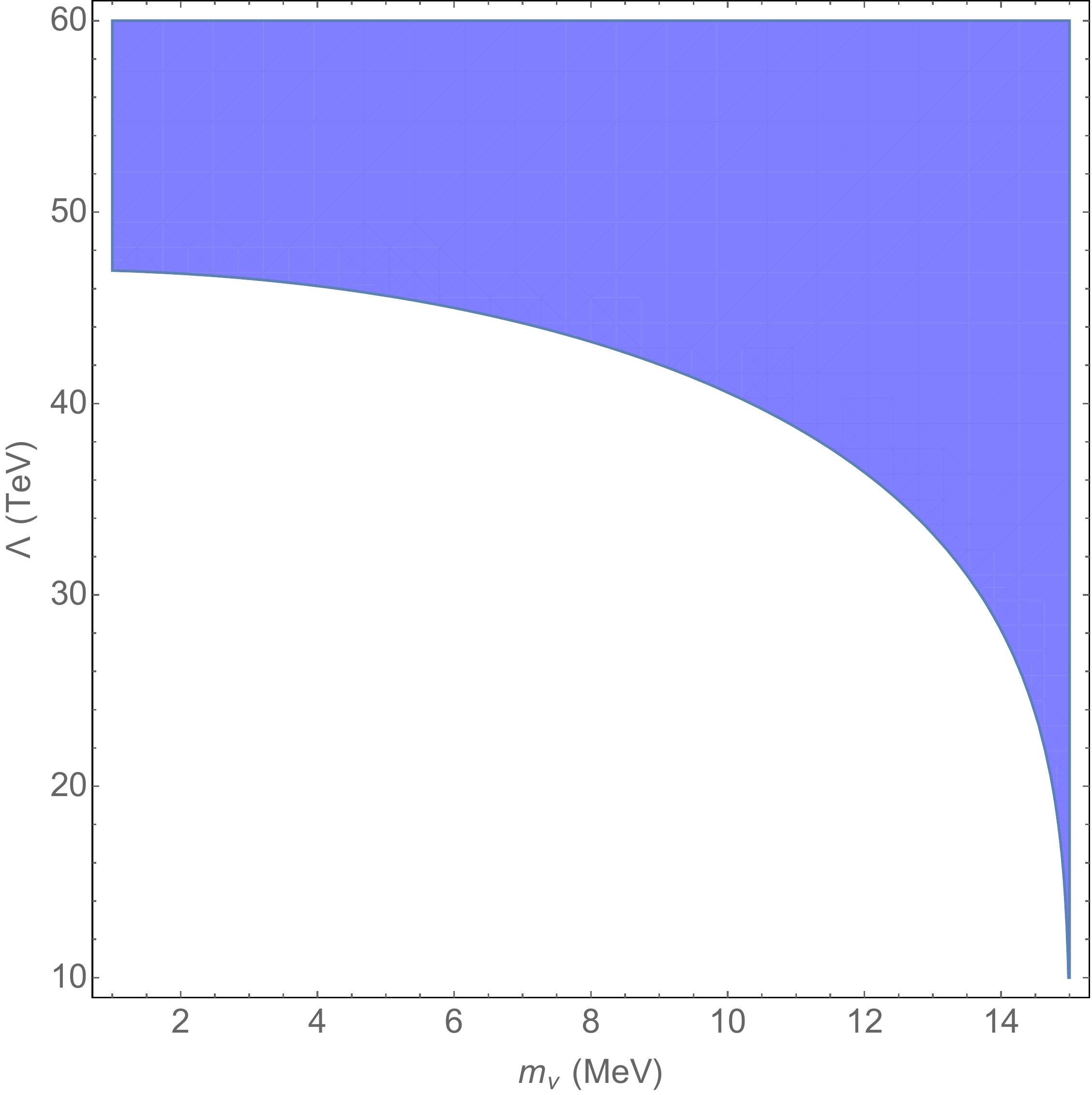}
$$
\caption{Shaded area: region in the neutrino mass $m_\nu$ (GeV) versus new physics scale $\Lambda$ (GeV) plane allowed by the neutrino magnetic moment constraint Eq.~(\ref{eq:mnubound}).}
\label{fig:NMM1}
\end{figure}

There are no important limits on $ \Lambda $ for  magnetic coupling for larger neutrino masses; in particular, collider data is useful only when the underlying physics is assumed to be strongly coupled. For weakly-coupled heavy physics the effective operator coefficients are too small to produce a measurable effect given the experimental sensitivity \cite{nuSM-5}.

\bigskip

When the above decays are forbidden, or when the operator does not have 3-legged vertices, the experimental constraints are degraded. This is because in this case existing limits are obtained within the context of specific models and cannot be directly extended to limits on all effective operator coefficients. For example, there are strict limits on the masses of the \rh\ neutrinos and the \rh\ $W_R$ gauge bosons present in L-R models \cite{Khachatryan:2014dka}. But these do not translate to limits on $m_\nu $ or $ \Lambda $ (when the $W_R$ and extra scalars are assumed heavy, with masses  $ \sim\Lambda$) since the leading effective operators (obtained after integration of these heavy particles) whose effects are strongly constrained by experiment, are {\em not} the ones being considered here. The simplest way of seeing this is to note that all our operators violate lepton number, so within such L-R models they will appear multiplied by small coefficients (e.g. by the small \vev\ of a scalar triplet~\cite{Dhuria:2015cfa}), and have subdominant effects. The point is that without knowledge of the model realized in Nature, one cannot, in general, use limits obtained using some operators to constrain the effects of others. A full analysis of the potential collider reactions that can best probe the operators in table~\ref{tab:rhn} lies beyond the scope of this paper.

\section{Conclusions}

In this paper we have worked out all possible effective operators of dimension 7 involving SM fields and right-handed neutrinos, indicating those that could be generated at tree generated by the underlying theory and which may then contribute significantly to low energy observables. Our results generalize lists of dimension 7 operators involving only SM fields \cite{dim7-1}, as well as earlier earlier partial compilations \cite{dim7-2}. All dimension 7 operators violate $B-L$ by two units.

From the operators presented, we selected two that can generate clear dilepton and trilpeton signatures at the LHC. The current limit on the scale of new physics is $ \simeq 330~ \gev$ (Eq. (\ref{eq:clim})) while the the sensitivity to the scale of new physics can reach $ \simeq 460 $ GeV (Eq. (\ref{eq:flim})), with the dilepton channel providing the highest sensitivity. We argue that these results correspond to sensitivity to the presence of heavy fermion triplets (which can also be responsible for type III seesaw mechanism for generating neutrino masses) and are competitive with the ones obtained form direct production. We argued that these limits are not necessarily superseded by the stronger ones derived from precision data given the inability to account for cancellations in the latter.

We have also obtained limits on the NP scale of operators containing right handed neutrinos from $Z,H,W$ decays, which are of the order of $\sim 500 ~{\rm GeV}$ or less for $Z,W$ semi leptonic decays; limits on the $H$ invisible decay  gives a stronger limit $\sim 800 ~{\rm GeV}$ (all with a dependence on the right-handed neutrino mass). As for the light neutrinos, much stronger constraints are obtained from the magnetic-moment effective operator, but only for small ($ < 30~ \mev $) masses.

Finally, we note that effective Lagrangian models have recently gained interest when studying the LHC sensitivity to various dark matter models; see, for example \cite{DM-EFT}.  

\section{Acknowledgements} The work of SB is partially funded by DST-INSPIRE grant no. PHY/P/SUB/01 at IIT Guwahati.

%\newpage
\appendix
\section{Appendix: Loop-generated and potentially tree generated operators}
\label{sec:lg.ptg}
In this appendix we present the arguments for determining whether an operator is necessarily generated by heavy physics loops (LG operators), or if there are types of new physics that can generate the operator at tree-level (potentially tree-generated or PTG operators). As throughout this paper we assume the NP is weakly coupled, and decoupling, and that the full theory is renormalizable. In  this case, denoting by $I,~E$ the number of internal and external lines respectively, by $L$ the number of loops and by $V_n$ the number of vertices with $n$ legs we have the well-known relations
\beq
L = I-V+1 \qquad \sum_{n\ge3} n V_n = 2I + E \qquad V = \sum_{n\ge3} V_n
\eeq
which for tree graphs ($L=0$) in renromaliable theories ($V_{n\ge5}=0$) imply
\beq
V_3 + 2V_4 = E-2
\eeq
When needed we will denote  heavy fermion by $ \Psi $ a heavy scalar by $ \Phi $ and a a heavy vector by $V$; correspondingly we denote SM vectors, fermions and scalars by $A$, $\psi$ and $ \varphi$ respectively. 

We will also need some basic properties of the couplings of matter and vector fields  in gauge theories. We denote by the index `$l$' a gauge direction associated with a light (SM) vector boson, while an index `$h$' will be associated with the heavy gauge-boson directions. Accordingly the generators are denoted $T^l$ and $T^h$ and the group structure constants take the generic form $ f_{lll, llh, lhh, hhh}$. The group generators in general connect light and heavy particle (fermion and scalar) directions, these we also denote by subindices $l$ and $h$:  $(T^h)_{ll}$, $(T^h)_{lh}$, $(T^h)_{hl}$, $(T^h)_{hh}$ and similarly for $T^l$. General properties of gauge theories imply that
\beq
f_{llh} = 0\,, \quad  (T^l)_{lh}= (T^l)_{hl}= 0
\label{eq:z1}
\eeq
Since a  \vev\ $ \vevof\Phi $ does not break the electroweak symmetry we also have
\beq
T^l \vevof\Phi =0 \,.
\label{eq:z2}
\eeq
Also, since $ T^h \vevof\Phi $ is a vector in the direction of a would-be Goldstone boson, it  follows that
\beq
(T^h\vevof\Phi)_l =0 
\label{eq:z3}
\eeq

We now use these relations to determine the LG or PTG character of the operators listed in section \ref{sec:d7_ops}.

\paragraph{$\psi^2 D^4$ operators.} %9 LG operators } 
These are of the form $ \nu^2 A_\mn A_{\rho\sigma},\,\nu^2 A_\mn \tilde A_{\rho\sigma} $, and it is straightforward to see that the possible tree diagrams involve a heavy vector, scalar or fermion. 
\bit
\item Heavy fermion exchange: the graphs have a $\nu \Psi A $ vertex and are then proportional to $(T^l)_{lh} = 0 $, because of Eq.~(\ref{eq:z1}).
\item Heavy vector exchange: the graphs have a $AAV$ vertex and are then proportional to $ f_{llh} =0 $, because of Eq.~(\ref{eq:z1}).
\item Heavy scalar exchange: the graphs have a $AA\Phi $ vertex, and are then proportional to $ \vevof\Phi\{T^l, T^{l'}\}\Phi' =0 $, because of Eq.~(\ref{eq:z2}).
\eit

\paragraph{$\psi^2 D^3 \vp$ operators.}% 8 LG operators }
These  are of the form $ \partial^\mu \psi \gamma^\nu \vp \psi A_\mn$ or $ \partial^\mu \psi \gamma^\nu \vp \psi \tilde{A}_\mn$ and the possible tree diagrams involve a heavy vector, scalar
or fermion. 
\bit
\item Heavy fermion exchange: the graphs have a $\psi \Psi A $ vertex and are then proportional to $(T^l)_{lh} = 0 $, because of Eq.~(\ref{eq:z1}).
\item Heavy vector exchange: the graphs have a $\vp A V$ vertex and are then proportional to $ \vevof\Phi\{T^l, T^h\}\vp =0 $, because of Eq.~(\ref{eq:z2}) and Eq.~(\ref{eq:z3}).
\item Heavy scalar exchange: the graphs have a $A\vp\Phi $ vertex, which is proportional to $ (T^l)_{hl} =0 $, because of Eq.~(\ref{eq:z1}).
\eit

\begin{figure}[h]
$$
\includegraphics[width=4in]{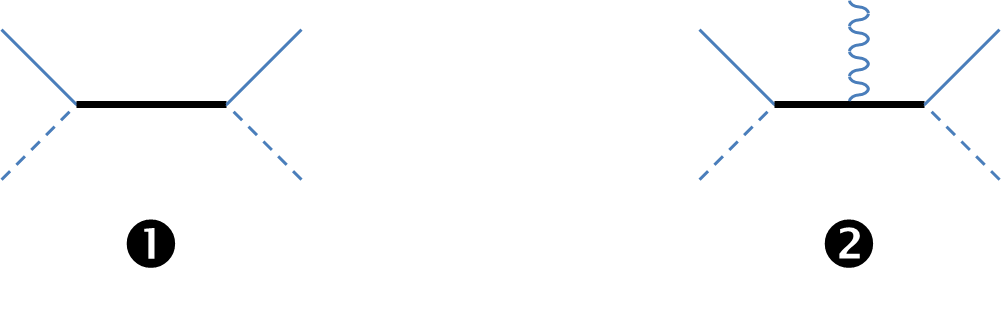}
$$
\caption{Diagrams that generate operators of the form $\psi^2 D^2 \vp^2$ at tree-level (thick lines denote particles with $(\Lambda)$ masses).}
\label{fig:p2D2}
\end{figure}

\paragraph{$\psi^2 D^2 \vp^2$ operators.}% 10 PTG operators }
These  separate in two categories according to whether $D^2$ represents a field tensor $ D^2 \to A_\mn$ or not. Both cases can be generated at tree level by the heavy-fermion exchange diagrams \ding{182} and \ding{183} in Fig.\ref{fig:p2D2}. None of the vertices in these graphs are {\it a priori} forbidden by the gauge symmetry (though they give rise to mixing among light and heavy fermions and the associated naturality issues).

There are three operators in this group, $\ocal_{1,\, 3,\, 4}$, that are associated with the type I, II and III seesaw neutrino mass generation mechanism. To see this relation explicitly we define  
\beq
\ocal'_I =(\ell\phi) \square (\ell\phi) \,,\quad
\ocal'_{II} = (\ell \taubf \ell) D^2 (\phi \taubf \phi)  \,, \quad
\ocal'_{III} = (\ell \taubf \phi) D^2 (\ell \taubf \phi)  \,,
\eeq 
which can be generated by the exchange of a fermion isosinglet, a scalar isotriplet or a fermion isotriplet, respectively; the relevant graphs are this in Fig. \ref{fig:p2D2}. Using the equations of motion (as allowed by the equivalence theorem, see section \ref{sec-0}) we find that these two set of operators are equivalent:
\beq
\bpm \ocal'_I \cr \ocal'_{II} \cr \ocal'_{III}  \epm = 
\bpm 
0  & 0 & 2 \cr
8  & 4 & 8 \cr
-2 & 0 &-6 \epm
\bpm \ocal_1 \cr \ocal_3 \cr \ocal_4  \epm
\eeq
The advantage of the primed basis is that it manifestly depicts the type of particles that generate them and nicely matches them to the usual see-saw graphs associated with neutrino mass generation.

Now, $\ocal_{1,3}$  contain the term $  \mw^2 W^+{}^2 e_L^2 $ ($ \ocal_2 $ does not), then $ \ocal'_{II,\,III} $ will also have such interactions.
The validity of the EFT for $ \ocal'_{II} $ requires the dilepton invariant mass satisfy $M_{\ell\ell} \ll \Lambda $, while for $ \ocal'_{III} $ ihe $W$-lepton should be similarly bound: $ M_{\ell W} \ll \Lambda $.

\paragraph{$\psi^2 D  \vp^3$ operators:}  contain vertices with 2 fermions and 3 scalars and it is easy to see that can be generated at tree level, for example by  $ \Psi $ exchange.

\paragraph{$\psi^2 \vp^4$ operators:} contain vertices with 2 fermions and 4 scalars,. The tree-level graphs then satisfy~\footnote{The case $V_4=2$ does not occur because in renormalizable theories there are no 4-legged vertices with fermions}  $ V_3 + 2 V_4 = 4$ so that $ V_3 =2,\, V_4=1$ or $ V_3=4$, and the first class of graphs can generate the 
operators at tree level (vertices: $ \psi^2 \Phi,\,\Phi^2\vp,\,\vp^3 \Phi $, and two internal $ \Phi$ lines).

\begin{figure}[h]
$$
\includegraphics[width=2in]{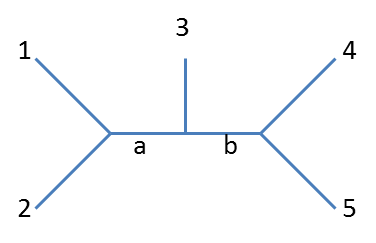}
$$
\caption{Tree-level graph that can generate operators of the form $ \psi^4 D$; lines labeled `a' and `b' correspond to heavy particles (see text for details).}
\label{fig:p4D}
\end{figure}

\paragraph{$\psi^4 D$ operators:} these contain a vertex $ \psi^4 A$, which corresponds to tree-level graphs with  $E=5$ and $ V_3=3,\,V_4 =0 $ ,or $ V_3=V_4=1 $, but the latter do not occur since the underlying (renormlizable) theory does not contain 4-legged vertices with fermions; graphs are those in Fig. \ref{fig:p4D}, where
\beq
\begin{array}{c|lllll|ll}
{\rm case} & 1 & 2 & 3 & 4 & 5 & a & b \cr
\hline
i   & \psi & \psi & A   & \psi  & \psi & \Phi & \Phi \cr  
ii  &\psi & \psi & A    & \psi  & \psi & V    & V    \cr  
iii &\psi & \psi & A    & \psi  & \psi & V    & \Phi \cr 
iv  & X    & \psi & \psi & \psi & \psi & \Psi &   V  \cr  
v   & X    & \psi & \psi & \psi & \psi & \Psi & \Phi \cr  
\end{array}
\label{eq:p4D}
\eeq
\bit
\item Case $i$ contains the vertex $\Phi\Phi X $ that has a derivative so the operators generated by such graphs are of the form $ \psi^4 X \partial $ and are of dimension 8.
\item Case $ii$ contains the vertex $VVA$ with one derivative and also corresponds to operators of dimension 8.
\item Case $iii$ contains the vertex $XV\Phi $ of the form which vanish because of  Eq.~(\ref{eq:z2}) and Eq.~(\ref{eq:z3}).
\item Cases $iv$ and $v$ require the vertex $ \psi X \Psi $ that does not exist because of Eq.~(\ref{eq:z1}).
\eit
If follows that all these are LG operators.

\paragraph{$\psi^4 \vp$ operators:} since 4-fermion interactions can be  tree-generated by exchange of a heavy boson, operators of the form $ \psi^4 \vp $ can be obtained by attaching a $ \varphi $ to the internal heavy boson line. These are all PTG operators.

\section{Appendix: PTG operators containing right handed neutrinos}
\label{sec:ptg.rh}

The dimension 7 PTG operators containing right-handed neutrinos (see section \ref{sec:d7_ops}) are listed in table \ref{tab:rhn}.
{\small
\begin{table}[ht]
%\begin{sidewaystable}[ht]
\hspace{-3in}
$$
\begin{array}{ccc}
\ocal~{\rm (see~sect~\ref{sec:d7_ops})} & \quad {\rm Unitary~gauge~expression} & \quad {\rm 3-legged~vertices~in~}\ocal{\rm~(unitary~gauge)}\cr \hline
\half (\ncb \gamma^\mu E )(\phi \epsilon \stackrel \leftrightarrow D_\mu \phi) &\quad - \frac i2 g  (v+H)^3 \overline{\nu^c} \slashed W^+ P_L e & - i \mw v^2 \overline{\nu^c} \slashed W^+ P_L e \cr
\half (\ncb \gamma^\mu N )(i \phi^\dagger \stackrel \leftrightarrow D_\mu \phi) &\quad  \inv{\sqrt{8}} ( H+v )^3 \overline{\nu^c} \slashed Z P_L \nu & \frac{v^3}{2\sqrt{2}}  \overline{\nu^c} \slashed Z P_L \nu\cr
 (\ncb \gamma^\mu N )(\partial_\mu |\phi|^2) &\quad - \inv{\sqrt{2}}  (H+v)^2 \partial_\mu H \left(\overline{\nu^c} \gamma^\mu P_L \nu \right) & - \frac{v^2 }{\sqrt{2}} \partial_\mu H \left(\overline{\nu^c} \gamma^\mu P_L \nu \right) \cr
 (\ncb \sigma^\mn e)(\phi\eps\tau^I \phi) W^I_\mn  &\quad \sqrt{2} (H+v)^2 \left[ \partial_\mu W_\nu^+ - i ( e A_\mu + g \cw Z_\mu) W_\nu^+ \right] \left( \overline{\nu^c} \sigma^\mn P_R e \right) & \frac{v^2}{\sqrt{2}} W_\mn^+ \left( \overline{\nu^c} \sigma^\mn P_R e \right)\cr 
 |\phi|^2 (\ncb \sigma^\mn \nu) B_\mn  &\quad  \half ( H+v)^2 \left( \overline{\nu^c} \sigma^\mn P_R \nu \right)\left( \cw F_\mn - \sw Z_\mn \right) & \half v^2 \left( \overline{\nu^c} \sigma^\mn P_R \nu \right)\left( \cw F_\mn - \sw Z_\mn \right) \cr
 (\ncb D_\mu e)(\phi\eps D^\mu \phi)  &\quad  \frac g{\sqrt{2}}  (v+H)^2 \left[ - i \left( \overline{\nu^c} \partial^\mu P_R e \right) W^+_\mu + \left( \overline{\nu^c} P_R e \right) W^+ \cdot(e A - g' \sw Z ) \right] & -i \sqrt{2}\, \mw v  W^+_\mu \left( \overline{\nu^c} \partial^\mu P_R e \right)\cr
  (\ncb \nu) |D\phi|^2  &\quad \half \left( \overline{\nu^c} P_R \nu \right) \left\{  \left[ \partial H + i \frac{\mz}v (H+v) Z \right]^2 + \half g^2 (v+H)^2 \mw^2 W^+ \cdot W^- \right\} & -- \cr
(\ncb\nu) |\phi|^4 &\quad  \inv4 ( H+v)^4 \left( \overline{\nu^c} P_R \nu \right) & v^3 H\left( \overline{\nu^c} P_R \nu \right)
\end{array}
$$
\caption{Dimension 7 operators containing \rh\ neutrinos, their unitary-gauge expressions, and the vertices with 3 fields that they contain. We defined $ F_\mn = \partial_\mu A_\nu - \partial_\nu A_\mu $, $ W^+_\mn = \partial_\mu W^+_\nu - \partial_\nu W^+_\mu $, and $ Z_\mn = \partial_\mu Z_\nu - \partial_\nu Z_\mu$;   $g,\,g'$ denote, respectively, the $\su2_L$ and $\ui_Y$ gauge coupling constants.}
\label{tab:rhn}
%\end{sidewaystable}
\end{table}
} % end of \small font

It follows from this list that there are 8 types of 3-legged vertices involving \rh\ neutrinos and contributing to the $Z,H,W$ decays and neutrino magnetic moments:
\beq
\begin{array}{|l|c|c|c|}
\hline
   & {\rm SM-like~coupling} & {\rm derivative~coupling} & {\rm magnetic~coupling}\cr \hline
Z & \overline{\nu^c} \slashed Z P_L \nu & -- & \left(\overline{\nu^c} \sigma^\mn P_R \nu \right)  Z_\mn\cr
H & H \left( \overline{\nu^c} P_R \nu \right) & \partial_\mu H \left(\overline{\nu^c} \gamma^\mu P_L \nu \right) & --\cr
W & \overline{\nu^c} \slashed W^+ P_L e & W^+_\mu \left( \overline{\nu^c} \partial^\mu P_R e \right) & \left( \overline{\nu^c} \sigma^\mn P_R e \right) W^+_\mn \cr
\gamma & -- & -- & \left(\overline{\nu^c} \sigma^\mn P_R \nu \right)  F_\mn \cr
\hline
\end{array}
\label{eq:3lv}
\eeq
The most significant constraints on $ \Lambda $ derived from these vertices were derived in section \ref{sec:rhn}.


\begin{thebibliography}{99}

\bibitem{neutrino-mass}

See for example, M.~C.~Gonzalez-Garcia and Y.~Nir,
  ``Neutrino masses and mixing: Evidence and implications,''
  Rev.\ Mod.\ Phys.\  {\bf 75}, 345 (2003)
  [hep-ph/0202058]; 
  R. Mohapatra, S. Antusch, K. Babu, G. Barenboim, M.-C. Chen, et al., "Theory of
neutrinos: A White paper," Rept.\ Prog.\ Phys. {\bf 70}, 1757 (2007), [arXiv:hep-ph/0510213].

\bibitem{dark-matter}

See for example, G.~Bertone, D.~Hooper and J.~Silk,
  ``Particle dark matter: Evidence, candidates and constraints,''
  Phys.\ Rept.\  {\bf 405}, 279 (2005)
  [hep-ph/0404175].
  
  \bibitem{eff-theory1}
  See for example, J.~Polchinski,
  ``Effective field theory and the Fermi surface,''
  In *Boulder 1992, Proceedings, Recent directions in particle theory* 235-274, and Calif. Univ. Santa Barbara - NSF-ITP-92-132 (92,rec.Nov.) 39 p. (220633) Texas Univ. Austin - UTTG-92-20 (92,rec.Nov.) 39 p [hep-th/9210046].
  
  \bibitem{eff-theory2}
  K.~G.~Wilson and J.~B.~Kogut,
  ``The Renormalization group and the epsilon expansion,''
  Phys.\ Rept.\  {\bf 12}, 75 (1974).
  
  \bibitem{strong1}
  S.~Weinberg,
  ``Phenomenological Lagrangians,''
  Physica A {\bf 96} (1979) 327;
  For a pedagogical review see: 
  A. Dobado, A. Gomez-Nicola, A. L. Maroto  and J. R. Pelaez, {\it Effective Lagrangians for the Standard Model} 
(Spinger, 1997; ISBN-10: 3540625704, ISBN-13: 978-3540625704)  
  
    
  % \bibitem{strong2}
  %N.~Brambilla, S.~Eidelman, P.~Foka, S.~Gardner, A.~S.~Kronfeld, M.~G.~Alford, R.~Alkofer and M.~Butenschoen {\it et al.},
  %``QCD and Strongly Coupled Gauge Theories: Challenges and Perspectives,''
 % Eur.\ Phys.\ J.\ C {\bf 74}, no. 10, 2981 (2014)
  %[arXiv:1404.3723 [hep-ph]].
  
   \bibitem{eqv-th}
  H.~Georgi,
  ``On-shell effective field theory,''
  Nucl.\ Phys.\ B {\bf 361} (1991) 339;
 J.~Wudka,
  ``Electroweak effective Lagrangians,''
  Int.\ J.\ Mod.\ Phys.\ A {\bf 9}, 2301 (1994)
  [hep-ph/9406205];
  C.~Arzt,
  ``Reduced effective Lagrangians,''
  Phys.\ Lett.\ B {\bf 342}, 189 (1995)
  [hep-ph/9304230].

  
   \bibitem{SM-renorm}
  G.~'t Hooft and M.~J.~G.~Veltman,
  ``Regularization and Renormalization of Gauge Fields,''
  Nucl.\ Phys.\ B {\bf 44}, 189 (1972).
  
  \bibitem{decoup}
  %\cite{Appelquist:1974tg}
%\bibitem{Appelquist:1974tg}
  T.~Appelquist and J.~Carazzone,
  ``Infrared Singularities and Massive Fields,''
  Phys.\ Rev.\ D {\bf 11} (1975) 2856.
  %%CITATION = PHRVA,D11,2856;%%
  For a pedagogical introduction see:  J.~Collins, {\it Renormalization: An Introduction to Renormalization, the Renormalization Group and the Operator-Product Expansion }
  (Cambridge University Press, 1986; ISBN-10: 0521311772, ISBN-13: 978-0521311779).
  
  
  \bibitem{Branco:2011iw} 
  G.~C.~Branco, P.~M.~Ferreira, L.~Lavoura, M.~N.~Rebelo, M.~Sher and J.~P.~Silva,
  ``Theory and phenomenology of two-Higgs-doublet models,''
  Phys.\ Rept.\  {\bf 516}, 1 (2012)
  [arXiv:1106.0034 [hep-ph]].
 
     
  \bibitem{Agashe:2004rs}
  K.~Agashe, R.~Contino and A.~Pomarol,
 ``The Minimal composite Higgs model,''
  Nucl.\ Phys.\ B {\bf 719} (2005) 165
  [hep-ph/0412089].
  %%CITATION = HEP-PH/0412089;%%
  
  \bibitem{Veltman:1980mj}
  M.~J.~G.~Veltman,
  ``The Infrared - Ultraviolet Connection,''
  Acta Phys.\ Polon.\ B {\bf 12} (1981) 437.
  %%CITATION = APPOA,B12,437;%%
  
  \bibitem{dim5}
  S.~Weinberg,
  ``Baryon and Lepton Nonconserving Processes,''
  Phys.\ Rev.\ Lett.\  {\bf 43}, 1566 (1979).
  
  \bibitem{dim6-1}
  W.~Buchmuller and D.~Wyler,
  ``Effective Lagrangian Analysis of New Interactions and Flavor Conservation,''
  Nucl.\ Phys.\ B {\bf 268}, 621 (1986).
  
  \bibitem{dim6-2}
  B.~Grzadkowski, M.~Iskrzynski, M.~Misiak and J.~Rosiek,
  ``Dimension-Six Terms in the Standard Model Lagrangian,''
  JHEP {\bf 1010}, 085 (2010)
  [arXiv:1008.4884 [hep-ph]].
  
      
   \bibitem{dim7-1}
  L.~Lehman,
  ``Extending the Standard Model Effective Field Theory with the Complete Set of Dimension-7 Operators,''
  Phys.\ Rev.\ D {\bf 90}, no. 12, 125023 (2014)
  [arXiv:1410.4193 [hep-ph]].
  
  \bibitem{dim8-9-10-11}
  K.~S.~Babu and C.~N.~Leung,
  ``Classification of effective neutrino mass operators,''
  Nucl.\ Phys.\ B {\bf 619}, 667 (2001)
  [hep-ph/0106054];
  A.~de Gouvea and J.~Jenkins,
  ``A Survey of Lepton Number Violation Via Effective Operators,''
  Phys.\ Rev.\ D {\bf 77}, 013008 (2008)
  [arXiv:0708.1344 [hep-ph]];
  F.~Bonnet, D.~Hernandez, T.~Ota and W.~Winter,
  ``Neutrino masses from higher than d=5 effective operators,''
  JHEP {\bf 0910}, 076 (2009)
  [arXiv:0907.3143 [hep-ph]];
%  F.~del Aguila, A.~Aparici, S.~Bhattacharya, A.~Santamaria and J.~Wudka,
%  ``Effective Lagrangian approach to neutrinoless double beta decay and neutrino masses,''
%  JHEP {\bf 1206}, 146 (2012)
%  [arXiv:1204.5986 [hep-ph]];
  P.~W.~Angel, N.~L.~Rodd and R.~R.~Volkas,
  ``Origin of neutrino masses at the LHC: $\Delta L = 2$ effective operators and their ultraviolet completions,''
  Phys.\ Rev.\ D {\bf 87}, no. 7, 073007 (2013)
  [arXiv:1212.6111 [hep-ph]];
  C.~Degrande,
  ``A basis of dimension-eight operators for anomalous neutral triple gauge boson interactions,''
  JHEP {\bf 1402}, 101 (2014)
  [arXiv:1308.6323 [hep-ph]].
  
   \bibitem{dim7-2}
   S.~Weinberg,
  ``Varieties of Baryon and Lepton Nonconservation,''
  Phys.\ Rev.\ D {\bf 22}, 1694 (1980);
  K.~S.~Babu and R.~N.~Mohapatra,
  ``B-L Violating Proton Decay Modes and New Baryogenesis Scenario in SO(10),''
  Phys.\ Rev.\ Lett.\  {\bf 109}, 091803 (2012)
  [arXiv:1207.5771 [hep-ph]];
  K.~S.~Babu and R.~N.~Mohapatra,
  ``B-L Violating Nucleon Decay and GUT Scale Baryogenesis in SO(10),''
  Phys.\ Rev.\ D {\bf 86}, 035018 (2012)
  [arXiv:1203.5544 [hep-ph]];
   G.~Chalons and F.~Domingo,
  ``Dimension 7 operators in the b to s transition,''
  Phys.\ Rev.\ D {\bf 89}, no. 3, 034004 (2014)
  [arXiv:1303.6515 [hep-ph]].

  
  \bibitem{eff-appl}
  
  H.~A.~Weldon and A.~Zee,
  ``Operator Analysis of New Physics,''
  Nucl.\ Phys.\ B {\bf 173}, 269 (1980);
  C.~Arzt, M.~B.~Einhorn and J.~Wudka,
  ``Effective Lagrangian approach to precision measurements: The Anomalous magnetic moment of the muon,''
  Phys.\ Rev.\ D {\bf 49}, 1370 (1994)
  [hep-ph/9304206].
  S.~Bar-Shalom, A.~Soni and J.~Wudka,
  ``EFT naturalness: an effective field theory analysis of Higgs naturalness,''
  arXiv:1405.2924 [hep-ph];
  S.~Dawson, I.~M.~Lewis and M.~Zeng,
  ``Effective field theory for Higgs boson plus jet production,''
  Phys.\ Rev.\ D {\bf 90}, no. 9, 093007 (2014)
  [arXiv:1409.6299 [hep-ph]];
  R.~Catena,
  ``Prospects for direct detection of dark matter in an effective theory approach,''
  JCAP {\bf 1407}, 055 (2014)
  [arXiv:1406.0524 [hep-ph]];
  S.~Matsumoto, S.~Mukhopadhyay and Y.~L.~S.~Tsai,
  ``Singlet Majorana fermion dark matter: a comprehensive analysis in effective field theory,''
  JHEP {\bf 1410}, 155 (2014)
  [arXiv:1407.1859 [hep-ph]];
  G.~M.~Pruna and A.~Signer,
  ``The $\mu\to e\gamma$ decay in a systematic effective field theory approach with dimension 6 operators,''
  JHEP {\bf 1410}, 14 (2014)
  [arXiv:1408.3565 [hep-ph]];
  A.~Biek�tter, A.~Knochel, M.~Kr�mer, D.~Liu and F.~Riva,
  ``Vices and virtues of Higgs effective field theories at large energy,''
  Phys.\ Rev.\ D {\bf 91}, 055029 (2015)
  [arXiv:1406.7320 [hep-ph]];
  M.~Duch, B.~Grzadkowski and J.~Wudka,
  ``Classification of effective operators for interactions between the Standard Model and dark matter,''
  JHEP {\bf 1505}, 116 (2015)
  [arXiv:1412.0520 [hep-ph]].
  
  \bibitem{majorana}
  
  See for example, A.~Zee,
  %``A Theory of Lepton Number Violation, Neutrino Majorana Mass, and Oscillation,''
  Phys.\ Lett.\ B {\bf 93}, 389 (1980)
  Erratum: [Phys.\ Lett.\ B {\bf 95}, 461 (1980)].
  
  
  \bibitem{nuSM-5}
  A.~Aparici, K.~Kim, A.~Santamaria and J.~Wudka,
  ``Right-handed neutrino magnetic moments,''
  Phys.\ Rev.\ D {\bf 80}, 013010 (2009)
  [arXiv:0904.3244 [hep-ph]].
  
  \bibitem{nuSM-6}
  %\cite{delAguila:2008ir}
%\bibitem{delAguila:2008ir}
  F.~del Aguila, S.~Bar-Shalom, A.~Soni and J.~Wudka,
  ``Heavy Majorana Neutrinos in the Effective Lagrangian Description: Application to Hadron Colliders,''
  Phys.\ Lett.\ B {\bf 670} (2009) 399
  [arXiv:0806.0876 [hep-ph]].
  %%CITATION = ARXIV:0806.0876;%%
  
  \bibitem{0nbb}
  
  See for example, J.~D.~Vergados,
  ``The Neutrinoless double beta decay from a modern perspective,''
  Phys.\ Rept.\  {\bf 361}, 1 (2002)
  [hep-ph/0209347];
  %\cite{delAguila:2012nu}
%\bibitem{delAguila:2012nu}
  F.~del Aguila, A.~Aparici, S.~Bhattacharya, A.~Santamaria and J.~Wudka,
  ``Effective Lagrangian approach to neutrinoless double beta decay and neutrino masses,''
  JHEP {\bf 1206} (2012) 146
  [arXiv:1204.5986 [hep-ph]].
  %%CITATION = ARXIV:1204.5986;%%
  
  \bibitem{Higgs-LHC}
  
 See for example, M.~Spira, A.~Djouadi, D.~Graudenz and P.~M.~Zerwas,
  ``Higgs boson production at the LHC,''
  Nucl.\ Phys.\ B {\bf 453}, 17 (1995)
  [hep-ph/9504378].
 %  
%\bibitem{Einhorn:2013tja}
  M.~B.~Einhorn and J.~Wudka,
  %``Higgs-Boson Couplings Beyond the Standard Model,''
  Nucl.\ Phys.\ B {\bf 877} (2013) 792
  [arXiv:1308.2255 [hep-ph]].
  %%CITATION = ARXIV:1308.2255;%%
    
  \bibitem{Jose-PTG}
  C.~Arzt, M.~B.~Einhorn and J.~Wudka,
  ``Patterns of deviation from the standard model,''
  Nucl.\ Phys.\ B {\bf 433}, 41 (1995)
  [hep-ph/9405214].
  
 
  
   \bibitem{Jose-basis}
  M.~B.~Einhorn and J.~Wudka,
  ``The Bases of Effective Field Theories,''
  Nucl.\ Phys.\ B {\bf 876}, 556 (2013)
  [arXiv:1307.0478 [hep-ph]]. 
  
  \bibitem{B-L}
  A.~de Gouvea, J.~Herrero-Garcia and A.~Kobach,
  ``Neutrino Masses, Grand Unification, and Baryon Number Violation,''
  Phys.\ Rev.\ D {\bf 90}, no. 1, 016011 (2014)
  [arXiv:1404.4057 [hep-ph]].

  \bibitem{Vergados:2012xy}
  J.~D.~Vergados, H.~Ejiri and F.~Simkovic,
  %``Theory of Neutrinoless Double Beta Decay,''
  Rept.\ Prog.\ Phys.\  {\bf 75} (2012) 106301
  doi:10.1088/0034-4885/75/10/106301
  [arXiv:1205.0649 [hep-ph]].
  
  \bibitem{delAguila:2012nu} See
  F.~del Aguila, A.~Aparici, S.~Bhattacharya, A.~Santamaria and J.~Wudka,
  %``Effective Lagrangian approach to neutrinoless double beta decay and neutrino masses,''
  JHEP {\bf 1206} (2012) 146
  doi:10.1007/JHEP06(2012)146
  [arXiv:1204.5986 [hep-ph]], and references therein.

  
  \bibitem{PDG}
  K.A. Olive et al. (Particle Data Group), Chin. Phys. C, 38, 090001 (2014).
  
  \bibitem{plasmon}
V.~Castellani and S.~Degl'Innocenti,
  %``Stellar evolution as a probe of neutrino properties,''
  Astrophys.\ J.\  {\bf 402} (1993) 574.
 M.~Catelan, J.~A.~d.~F.~Pacheco and J.~E.~Horvath,
  %``The helium-core mass at the helium flash in low-mass red giant stars: observations and theory,''
  Astrophys.\ J.\  {\bf 461} (1996) 231
  [astro-ph/9509062].
   M.~Haft, G.~Raffelt and A.~Weiss,
  %``Standard and nonstandard plasma neutrino emission revisited,''
  Astrophys.\ J.\  {\bf 425} (1994) 222
   [Astrophys.\ J.\  {\bf 438} (1995) 1017]
  [astro-ph/9309014].
   G.~G.~Raffelt,
  %``Core Mass at the Helium Flash From Observations and a New Bound on Neutrino Electromagnetic Properties,''
  Astrophys.\ J.\  {\bf 365} (1990) 559.
  G.~G.~Raffelt,
  %``New bound on neutrino dipole moments from globular cluster stars,''
  Phys.\ Rev.\ Lett.\  {\bf 64} (1990) 2856.
   G.~Raffelt and A.~Weiss,
  %``Nonstandard neutrino interactions and the evolution of red giants,''
  Astron.\ Astrophys.\  {\bf 264} (1992) 536.
  A.~Heger, A.~Friedland, M.~Giannotti and V.~Cirigliano,
  %``The Impact of Neutrino Magnetic Moments on the Evolution of Massive Stars,''
  Astrophys.\ J.\  {\bf 696} (2009) 608
  [arXiv:0809.4703 [astro-ph]].A.~Heger, A.~Friedland, M.~Giannotti and V.~Cirigliano,
  %``The Impact of Neutrino Magnetic Moments on the Evolution of Massive Stars,''
  Astrophys.\ J.\  {\bf 696} (2009) 608
  [arXiv:0809.4703 [astro-ph]].
  G. Raffelt, {\it Stars as Laboratories for Fundamental Physics: The Astrophysics of Neutrinos, Axions, and Other Weakly Interacting Particles}, (U. Chicago Press, 1996; ISBN-10: 0226702723, ISBN-13: 978-0226702728)     
  N.~Iwamoto, L.~Qin, M.~Fukugita and S.~Tsuruta,
  %``Neutrino magnetic moment and neutron star cooling,''
  Phys.\ Rev.\ D {\bf 51} (1995) 348.
  G.~G.~Raffelt,
  %``Limits on neutrino electromagnetic properties: An update,''
  Phys.\ Rept.\  {\bf 320} (1999) 319.

\bibitem{delAguila:2011gr}
  F.~del Aguila, A.~Aparici, S.~Bhattacharya, A.~Santamaria and J.~Wudka,
  %``A realistic model of neutrino masses with a large neutrinoless double beta decay rate,''
  JHEP {\bf 1205} (2012) 133
  doi:10.1007/JHEP05(2012)133
  [arXiv:1111.6960 [hep-ph]].


\bibitem {CalcHEP}
A.~Belyaev, N.~D.~Christensen and A.~Pukhov,
  ``CalcHEP 3.4 for collider physics within and beyond the Standard Model,''
  Comput.\ Phys.\ Commun.\  {\bf 184} (2013) 1729
  [arXiv:1207.6082 [hep-ph]].

\bibitem{CTEQ}
H.~L.~Lai {\it et al.}  [CTEQ Collaboration],
  ``Global {QCD} analysis of parton structure of the nucleon: CTEQ5 parton
  distributions,''
  Eur.\ Phys.\ J.\  C {\bf 12}, 375 (2000)
  [arXiv:hep-ph/9903282].

  \bibitem{ATLAS1}
  [ATLAS Collaboration],
  ``Physics at a High-Luminosity LHC with ATLAS,''
  arXiv:1307.7292 [hep-ex].
  
  \bibitem{CMS1}
  J.~Varela [CMS Collaboration],
  ``Prospects for physics at high luminosity with CMS,''
  EPJ Web Conf.\  {\bf 49}, 11003 (2013).

\bibitem{HL-LHC}
ATLAS Collaboration, `Physics at a High-Luminosity LHC with ATLAS',
ATL-PHYS-PUB-2012-001, https://cds.cern.ch/record/1472518.

\bibitem{topmass} The ATLAS, CDF, CMS and D0 Collaborations, First combination of Tevatron and LHC measurements of the top-quark mass; arXiv:1403.4427 [hep-ex].

 \bibitem{Pythia}
T.~Sjostrand, S.~Mrenna and P.~Skands,
  ``PYTHIA 6.4 physics and manual,''
  JHEP {\bf 0605}, 026 (2006)
  [arXiv:hep-ph/0603175].

\bibitem {ttbar-7}
V.~Khachatryan {\it et al.}  [CMS Collaboration],
  ``First Measurement of the Cross Section for Top-Quark Pair Production in Proton-Proton Collisions at $\sqrt{s}=7$ TeV,''
  Phys.\ Lett.\ B {\bf 695} (2011) 424
  [arXiv:1010.5994 [hep-ex]].
  
\bibitem{ttbar-8}
A.~Calderon [ATLAS and CMS Collaborations],
  ``Top pair cross section measurements at the LHC,''
  arXiv:1301.1158 [hep-ex].

\bibitem {ttbar-14}
S.~Moch and P.~Uwer,
  ``Theoretical status and prospects for top-quark pair production at hadron
  colliders,''
  Phys.\ Rev.\  D {\bf 78} (2008) 034003
  [arXiv:0804.1476 [hep-ph]].
  
  \bibitem{CMS-8TeV}
S.~Chatrchyan {\it et al.}  [CMS Collaboration],
  ``Search for new physics in events with same-sign dileptons and jets in pp collisions at $\sqrt{s}$ = 8 TeV,''
  JHEP {\bf 1401} (2014) 163
  [arXiv:1311.6736, arXiv:1311.6736 [hep-ex]].
  
  \bibitem{type3}
  R.~Foot, H.~Lew, X.~G.~He and G.~C.~Joshi,
  ``Seesaw Neutrino Masses Induced by a Triplet of Leptons,''
  Z.\ Phys.\ C {\bf 44}, 441 (1989).
  
  \bibitem{type3iv}
CMS Physics Analysis Summary, CMS PAS EXO-11-073.

\bibitem{CMSdimuon}
V.~Khachatryan {\it et al.} [CMS Collaboration],
  ``Search for heavy Majorana neutrinos in $\mu^\pm \mu^\pm+$ jets events in proton-proton collisions at $\sqrt{s}$ = 8 TeV,''
  Phys.\ Lett.\ B {\bf 748}, 144 (2015)
  doi:10.1016/j.physletb.2015.06.070
  [arXiv:1501.05566 [hep-ex]].
  
  \bibitem{type3i}
  F.~del Aguila and J.~A.~Aguilar-Saavedra,
  `Distinguishing seesaw models at LHC with multi-lepton signals,''
  Nucl.\ Phys.\ B {\bf 813}, 22 (2009)
  [arXiv:0808.2468 [hep-ph]].
    
  \bibitem{type3ii}
  R.~Franceschini, T.~Hambye and A.~Strumia,
  ``Type-III see-saw at LHC,''
  Phys.\ Rev.\ D {\bf 78}, 033002 (2008)
  [arXiv:0805.1613 [hep-ph]].
 
  
  \bibitem{type3iii}
  B.~Bajc and G.~Senjanovic,
  ``Seesaw at LHC,''
  JHEP {\bf 0708}, 014 (2007)
  [hep-ph/0612029].
  
%
\bibitem{delAguila:2008ir}
See, for example:
  F.~del Aguila, S.~Bar-Shalom, A.~Soni and J.~Wudka,
  %``Heavy Majorana Neutrinos in the Effective Lagrangian Description: Application to Hadron Colliders,''
  Phys.\ Lett.\ B {\bf 670} (2009) 399
  doi:10.1016/j.physletb.2008.11.031
  [arXiv:0806.0876 [hep-ph]].

%
\bibitem{Khachatryan:2014dka}
  V.~Khachatryan {\it et al.} [CMS Collaboration],
  %``Search for heavy neutrinos and $\mathrm {W}$ bosons with right-handed couplings in proton-proton collisions at $\sqrt{s} = 8\,\text {TeV} $,''
  Eur.\ Phys.\ J.\ C {\bf 74} (2014) 11,  3149
  doi:10.1140/epjc/s10052-014-3149-z
  [arXiv:1407.3683 [hep-ex]].
  
  %
\bibitem{Dhuria:2015cfa}
  M.~Dhuria, C.~Hati, R.~Rangarajan and U.~Sarkar,
  %``Falsifying leptogenesis for a TeV scale $W^{\pm}_{R}$ at the LHC,''
  Phys.\ Rev.\ D {\bf 92} (2015) 3,  031701
  doi:10.1103/PhysRevD.92.031701
  [arXiv:1503.07198 [hep-ph]].

 \bibitem{invisible_higgs2}
  G.~Aad {\it et al.} [ATLAS Collaboration],
  %``Search for invisible decays of a Higgs boson using vector-boson fusion in $pp$ collisions at $\sqrt{s}=8$ TeV with the ATLAS detector,''
  JHEP {\bf 1601}, 172 (2016)
  doi:10.1007/JHEP01(2016)172
  [arXiv:1508.07869 [hep-ex]].

\bibitem{DM-EFT}
J.~Goodman, M.~Ibe, A.~Rajaraman, W.~Shepherd, T.~M.~P.~Tait and H.~B.~Yu,
  ``Constraints on Dark Matter from Colliders,''
  Phys.\ Rev.\ D {\bf 82}, 116010 (2010)
  [arXiv:1008.1783 [hep-ph]];
  G.~Busoni, A.~De Simone, J.~Gramling, E.~Morgante and A.~Riotto,
  ``On the Validity of the Effective Field Theory for Dark Matter Searches at the LHC, Part II: Complete Analysis for the $s$-channel,''
  JCAP {\bf 1406}, 060 (2014)
  [arXiv:1402.1275 [hep-ph]];
G.~Busoni, A.~De Simone, T.~Jacques, E.~Morgante and A.~Riotto,
  ``On the Validity of the Effective Field Theory for Dark Matter Searches at the LHC Part III: Analysis for the $t$-channel,''
  JCAP {\bf 1409}, 022 (2014)
  [arXiv:1405.3101 [hep-ph]];
  J.~Abdallah, A.~Ashkenazi, A.~Boveia, G.~Busoni, A.~De Simone, C.~Doglioni, A.~Efrati and E.~Etzion {\it et al.},
  ``Simplified Models for Dark Matter and Missing Energy Searches at the LHC,''
  arXiv:1409.2893 [hep-ph].
 %
 

\end{thebibliography}
\end{document}